\pdfoutput=1
\documentclass[11pt]{article}
\usepackage{jheppub}
\usepackage{subfigure}

\def\matrixone{\hbox{$1\hskip -1.2pt\vrule depth 0pt height 1.6ex width 0.7pt
\vrule depth 0pt height 0.3pt width 0.12em$}}
\newcommand{\beq}{\begin{eqnarray}}
\newcommand{\eeq}{\end{eqnarray}}

\title{Fundamental Composite Electroweak Dynamics: Status at the LHC }
\author[a,b,c]{Alexandre Arbey,}
\author[d]{Giacomo Cacciapaglia,}
\author[d]{Haiying Cai,}
\author[d,1]{Aldo Deandrea,\note{also Institut Universitaire de France, 103 boulevard Saint-Michel, 75005 Paris, France}}
\author[b]{Sol\`ene Le Corre,}
\author[e]{Francesco Sannino}
\affiliation[a]{Universit{\' e} de Lyon, Universit{\' e} Lyon 1, F-69622 Villeurbanne Cedex, France; 
Centre de Recherche Astrophysique de Lyon, CNRS, UMR 5574, Saint-Genis-Laval Cedex, F-69561, France.}
\affiliation[b]{Ecole Normale Sup\'erieure
de Lyon, 46, all\'ee d'Italie, F-69364 Lyon cedex 07, France.}
\affiliation[c]{Theory Division, CERN, CH-1211 Geneva 23, Switzerland.}
\affiliation[d]{Universit\'e de Lyon, France; Universit\'e Lyon 1,
CNRS/IN2P3, UMR5822 IPNL,\\ F-69622 Villeurbanne Cedex, France.}
\affiliation[e]{CP$^3$-Origins and the Danish Institute for
Advanced Study DIAS, University of Southern Denmark, Campusvej
55, DK-5230 Odense M, Denmark.} 
\emailAdd{alexandre.arbey@ens-lyon.fr} 
\emailAdd{g.cacciapaglia@ipnl.in2p3.fr}
\emailAdd{deandrea@ipnl.in2p3.fr} 
\emailAdd{hcai@ipnl.in2p3.fr}
\emailAdd{solene.lecorre@ens-lyon.fr}
\emailAdd{sannino@cp3-origins.net}
\abstract{We determine the current status of the fundamental composite electroweak dynamics paradigm after the discovery of the Higgs boson at the Large Hadron Collider experiments. Our analysis serves as universal and minimal template for a wide class of models with the two limits in parameter space being composite Goldstone Higgs models and Technicolor.
This is possible  because of the existence of a unified description, both at the effective and fundamental Lagrangian levels, of models of composite Higgs dynamics where the Higgs boson itself can emerge, depending on the way the electroweak symmetry is embedded, either as a pseudo-Goldstone boson or as a massive excitation of the condensate. We constrain the available parameter space at the effective Lagrangian level. We show that a wide class of models of fundamental composite electroweak dynamics, including Technicolor, are compatible with experiments. The results are relevant for future searches of a fundamental composite nature of the Higgs mechanism at the Large Hadron Collider.} 
\preprint{CERN-PH-TH/2015-027 \\[-0.7cm]
\begin{flushright}
{LYCEN 2015-02} \\
{CP3-Origins-2015-007} 
 \end{flushright}}
\begin{document}
\maketitle

\section{Introduction}
The discovery of the Higgs boson is a landmark that establishes on a firmer experimental ground the Standard Model (SM) of particle physics. More excitingly, however, this discovery constitutes an invaluable source of information to help unveiling a more fundamental theory of particle interactions. The SM, in fact, suffers from a number of theoretical and phenomenological shortcomings such as the absence of a mechanism stabilising the electroweak scale against quantum corrections or of a dark matter candidate. For these reasons the SM can be seen as an effective description in search of a more fundamental one.

Despite the fact that a fully satisfactory underlying theory has yet  to be found, it is however possible to use the new experimental information on the Higgs sector to constrain extensions of the SM, which render this sector at least more fundamental.  
We shall focus on the possibility that the Higgs sector of the SM is composed of a new fundamental strongly coupled dynamics. The Higgs particle could then naturally emerge in two ways: mostly as a pseudo-Nambu Goldstone Boson (pNGB) \cite{Kaplan:1983fs,Kaplan:1983sm};  or mostly as the first composite scalar fluctuation of the new fundamental fermion condensate of Technicolor (TC) inspired theories. In general, it will be a linear combination of both states. 

Although within an effective Lagrangian description these different realisations seem superficially different, in fact,  at a more fundamental level  one can show that  any underlying four-dimensional composite pNGB nature of the Higgs is always accompanied by the TC limit. These two deceivingly different phenomenological realisations of the Higgs are, de facto, unified at the fundamental level \cite{Cacciapaglia:2014uja}. They  differ only in the final dynamical alignment of the electroweak symmetry and its embedding in the larger global symmetry of the fundamental theory. The converse is not true, i.e. one can have fundamental theories breaking the electroweak (EW) symmetry dynamically without admitting the pNGB Higgs limit.

The time-honoured example of fundamental descriptions of composite Higgs theories is Technicolor \cite{Weinberg:1975gm,Susskind:1978ms} where a scaled-up QCD dynamics was employed.  The original Weinberg and Susskind TC models, unfortunately, suffer of a number of serious phenomenological shortcomings and are, therefore, not phenomenologically viable. Among these issues there is the fact that the lightest massive composite scalar of the theory, the $\sigma(600)$,  when scaled up to the electroweak scale has a mass of around 1.5~TeV that can hardly be reconciled with experiments \cite{Sannino:2009za}. Constraints on TC models coming from the flavour sector must be taken {\it cum  grano salis} because they assume knowledge of extra, yet unspecified, sectors needed to endow the SM  fermions with mass. The interplay of these sectors with the one responsible for breaking the electroweak symmetry typically modifies the constraints \cite{Fukano:2009zm,Fukano:2010yv,Evans:2010ed,Foadi:2012bb,Foadi:2012ga,DiChiara:2014uwa}. The issue of flavour has also been analyses in the context of extra dimensional set-ups (see for instance~\cite{Fitzpatrick:2007sa,Albrecht:2009xr}), which however cannot be considered on the same footing as fundamental theories~\cite{Parolini:2014rza}.

 These shortcomings are not generalisable to other fundamental models of composite dynamics  \cite{Sannino:2009za}.    There is, in fact, a vast number of possible underlying theories at our disposal \cite{Sannino:2004qp,Dietrich:2006cm,Sannino:2009aw} that can be used to break the electroweak symmetry dynamically. For these theories the phenomenological constraints of Weinberg and Susskind TC models do not automatically apply, the reason being that the resulting composite dynamics can be very different from QCD. In particular modern models of fundamental dynamical electroweak symmetry breaking are based on the use of both different gauge groups and different underlying fermionic matter representations, as summarised in \cite{Sannino:2009za}.  It is therefore important to test  these new fundamental theories against data, especially because these models have the ambition to use a more fundamental nature of the Higgs boson and its sector. Similarly the composite Higgs of pNGB nature, if realised in nature beyond an effective description, should also be related to an underlying composite dynamics.

Following the results of \cite{Cacciapaglia:2014uja} we wish to determine the experimental status, via an effective Lagrangian approach, of the 
scalar sector of theories unifying the composite pNGB and  Techni-Higgs  at a more fundamental level.  The physical 125 GeV Higgs boson is 
therefore identified with the lightest state of the theory  which is generically a mixture of a composite pNGB  and the Techni-Higgs state. 

Although we are not using QCD as a template for our model building it is a fact that it contains in its spectrum a plethora of composite states, i.e. 
pNGBs, massive (pseudo)scalar resonances, axial and vector states, baryons (composite fermions) as well as high spin states. Even if the composite 
dynamics is not QCD-like a {\it smoking gun} evidence that new fundamental composite dynamics drives electroweak symmetry breaking would be 
the discovery of new composite states.  
 
Typically the phenomenology of non-perturbative extensions of the SM is limited to the bottom-up approach that lacks, however, of  specific predictions, for example for the actual spectrum of particles to be discovered, relevant to guide experimental searches. One would, in fact, like to have realistic expectations on when new states will be discovered at colliders.  In ref. \cite{Cacciapaglia:2014uja,Ryttov:2008xe,Galloway:2010bp}  a minimal realisation  in terms of an underlying gauge theory was provided consisting in a new underlying Fundamental Composite Dynamics (FCD), i.e. SU(2)$_{\rm FCD}$ gauge theory with two Dirac fundamental fermions transforming according to the defining representation of the gauge group. The non-perturbative chiral dynamics of this theory is being studied via first principle lattice simulations with noteworthy results. We know now, for example, that the pattern of chiral symmetry breaking that we shall be using below, i.e.  SU(4) to Sp(4), occurs dynamically  \cite{Lewis:2011zb,Hietanen:2013fya,Hietanen:2014xca}. Recently the spectrum of the lightest spin-one states appeared in  \cite{Hietanen:2013fya,Hietanen:2014xca}, while preliminary results for the scattering lengths and lattice signals  of a scalar state appeared in \cite{Arthur:2014zda,Arthur:2014lma}.  Equally important lattice results for the spectrum of minimal fundamental models of dynamical electroweak symmetry breaking that do not admit a composite pNGB Higgs limit are being produced with fermions in the adjoint representation  \cite{Catterall:2007yx,Catterall:2008qk,DelDebbio:2008wb,DelDebbio:2008zf,Catterall:2009sb,Hietanen:2009az,DelDebbio:2009fd,Kogut:2010cz,Karavirta:2011zg,Lewis:2011zb,Hietanen:2012qd,Hietanen:2012sz,Hietanen:2013fya,Hietanen:2013gva}. Direct experimental searches for these models have recently appeared in \cite{Aad:2014cka}.  

We define by {\it Fundamental Composite Electroweak Dynamics} any  four dimensional physical realisation of the Higgs sector of the SM in terms of underlying asymptotically free gauge theories capable of generating dynamically electroweak symmetry breaking. From the results of \cite{Cacciapaglia:2014uja} it is clear that pNGB Higgs, near-conformal as well as traditional TC theories are special limits of this universal definition that encompasses them all. 

Since the theory parameter space of well defined fundamental asymptotically free gauge theories is vast \cite{Sannino:2009za}, and given that some of these theories can describe simultaneously several phenomenologically viable extensions of the SM  \cite{Cacciapaglia:2014uja}  it is clear that fundamental composite dynamic theories are still prime candidates to be searched for at colliders.   

The paper is structured as follows: in Section~\ref{sec:model} we recap the main features of the model. The bounds from electroweak precision measurements and the Higgs couplings are presented in Section~\ref{sec:ewpt}, while in Section~\ref{sec:eta} we discuss the prospects to observe the lightest new particle, 
the singlet $\eta$, at the LHC and at a Linear Collider. Finally, in Section~\ref{sec:h2} we show the experimental bounds on the heavier Higgs, before concluding in Section~\ref{sec:concl}.

\section{The model} \label{sec:model}

In this paper we focus on a unified and minimal description of composite pNGB Higgs and TC models stemming from the simplest realisation in terms of an underlying fundamental dynamics.
Here, by simplest, we mean that it is based on the smallest asymptotically free gauge group with the smallest number of fermions needed for model building.  
The model~\cite{Cacciapaglia:2014uja,Ryttov:2008xe,Galloway:2010bp} relies on a gauge SU(2)$_{\rm FCD}$ strongly coupled group with just 
two Dirac fermions transforming according to the fundamental representation of the underlying gauge group~\footnote{This model was first proposed as a UV completion of Little Higgs models in~\cite{Batra:2007iz}.}. Since the representation is pseudo-real, 
the new fermions can be described as 4 Weyl fermions $Q^i$, so that the global symmetry of the fermionic sector is SU(4). The additional classical 
U(1) global symmetry is anomalous at the quantum level \footnote{The physical consequences are interlaced with the possible addition of the 
topological gauge-term \cite{DiVecchia:2013swa}.}.  Because SU(2) can be viewed as the first of the symplectic groups \cite{Sannino:2009aw}  the 
phenomenological analysis, and model building can be generalised to Sp($2N$)$_{\rm FCD}$~\cite{Barnard:2013zea}. 

The underlying Lagrangian is:
\begin{equation}
 \mathcal{L}=-\frac{1}{4}F^{a}_{\mu\nu}F^{a\mu\nu}+\bar{Q}_j (i\sigma^{\mu}D_{\mu})Q_j - M^{ij}_Q Q_i Q_j + h.c.
\end{equation}
with $F^{a}_{\mu\nu}$ the field strength of the FCD group, and $M_Q$ is a general mass matrix.
First principle numerical simulations~\cite{Lewis:2011zb,Hietanen:2013fya,Hietanen:2014xca}  have demonstrated that the SU(2)$_{\rm FCD}$ model does lead to a fermion condensate in the chiral limit breaking the global symmetry SU(4)$\to$ Sp(4). The group-theoretical properties of the condensate are: 
\begin{equation}
\langle Q^i Q^j \rangle = {\bf 6}_{\rm SU(4)} \to {\bf 5}_{\rm Sp(4)} \oplus {\bf 1}_{\rm Sp(4)}\,,
\end{equation}
transforming as a 2-index anti-symmetric representation of SU(4).
The coset space SU(4)/Sp(4) is parametrised by 5 Goldstone bosons, transforming as a $\bf 5$ of Sp(4)~\cite{Katz:2005au,Gripaios:2009pe}.

We need now to specify the embedding of the electroweak symmetry in the model: the simplest choice is to assign the first two $Q^i$ to a doublet of SU(2)$_L$, and the second two to a doublet of SU(2)$_R$ (the diagonal generator of SU(2)$_R$ being the generator of hypercharge). In this way, all gauge anomalies vanish, and we can keep track explicitly of the custodial symmetry built in the model.
Another point is that with this embedding, we can choose an alignment of the condensate in SU(4) that does not break the EW symmetry: this direction is in fact determined by the mass matrix $M_Q$.
The most general gauge-invariant mass term can be written as:
\begin{equation} \label{eq:MQ}
M_Q = \left( \begin{array}{cc}
\mu_L\; i \sigma_2 & 0 \\
0 & \mu_R\; i \sigma_2
\end{array} \right)\,,
\end{equation}
where $\sigma_2$ is the second Pauli matrix, and the phases of the techni-quarks can be used to make the two parameters $\mu_{L/R}$ real.
This mass term explicitly breaks SU(4) to Sp(4) in the case where $|\mu_L| = |\mu_R|$. In the following, we will choose $\mu_R = - \mu_L$ in order to use the same alignment of the vacuum as in \cite{Cacciapaglia:2014uja}, however the sign choice is arbitrary and irrelevant as it can always be reversed by a change in the phase of the constituent quarks.
All the physical results, therefore, are independent on the phases appearing in the mass matrix and in the condensate, provided we do not include the topological term  \cite{DiVecchia:2013swa}. 
The EW preserving vacuum, aligned with the mass matrix, is therefore
\begin{equation}
\Sigma_B = \left( \begin{array}{cc}
i \sigma_2 & 0 \\
0 & - i \sigma_2
\end{array} \right)\,.
\end{equation}
A list of the 10 unbroken generators $S^i$ and of the 5 broken ones $X^j$ can be found in Ref.  \cite{Cacciapaglia:2014uja}.
In this basis, the SU(2)$_L$ generators are $S^{1,2,3}$, while the SU(2)$_R$ ones are $S^{4,5,6}$.
The alignment of the condensate can be changed by applying an SU(4) transformation generated by the 5 broken generators: as $X^{1,2,3,4}$ form an SU(2) doublet, one can use gauge transformations to align the vacuum along the Higgs direction ($X^4$ in our notation) without loss of generality.
On the other hand, $X^5$ corresponds to a singlet of the gauged subgroup of SU(4), therefore a rotation along this direction will not change the physics of the model.
Furthermore, it can be shown that a transformation $e^{i \theta' X^5}$ will generate a relative phase between the mass terms $\mu_L$ and $\mu_R$ of the two techni-quark doublets: as already explained, this phase is irrelevant and can always be removed by a phase redefinition of the quarks. In other words, our choice to have real masses already fixed $\theta' = 0$. Introducing $\theta'$ in the vacuum alignment will therefore not add any new physical effects in the theory.
The most general vacuum, therefore, can be written as:
\begin{equation}
\Sigma_0 = e^{i \gamma} e^{i \theta X^4} \cdot \Sigma_B = e^{i \gamma} \left( \begin{array}{cc}
\cos \theta\; i \sigma_2 & \sin \theta \; \matrixone_{2\times 2} \\
- \sin \theta\; \matrixone_{2\times 2} & -\cos \theta\; i \sigma_2
\end{array} \right)\,.
\end{equation}
The phase $\gamma$ is generated by the anomalous U(1) symmetry, and it may therefore carry physical effects: in fact, it will generate CP violation in the chiral Lagrangian via the Pfaffian of the pion matrix~\cite{Galloway:2010bp}. In the following, for simplicity, we will limit ourselves to a CP-invariant model, thus setting $\gamma = 0$.
The only free parameter $\theta$ aligns the condensate to a direction that does break the EW symmetry, and its value will be determined once quantum corrections are added.

Based on the above symmetry considerations, one can describe the physics of the 5 Goldstone bosons via the CCWZ formalism~\cite{Galloway:2010bp,Katz:2005au,Gripaios:2009pe}: here we will use a linearly transforming matrix defined as
\begin{equation}
\Sigma = e^{i \sum_{j=1}^5 Y^j \chi_j/f} \cdot \Sigma_0\,,
\end{equation}
where $\chi_j$ are the pNGB fields, and $Y^j = e^{i \theta/2 X^4} \cdot X^j \cdot e^{-i \theta/2 X^4}$ are the broken generators in the $\Sigma_0$ vacuum.
For our purposes, this formalism is completely equivalent to the one based on 1-forms.
The chiral Lagrangian is therefore given by:
\begin{eqnarray}
\mathcal{L}_{\rm CCWZ} &=& \kappa_G (\sigma) f^2 \mbox{Tr} [(D_\mu \Sigma)^\dagger D^\mu \Sigma] + \frac{1}{2} \partial_\mu \sigma \partial^\mu \sigma - \frac{1}{2} M^2 \kappa_M (\sigma) \sigma^2  + \label{eq:CCWZ}\\
& & f \left( \kappa_t (\sigma)\, {y'}^{ij}_u (Q_{L,i} u_{R,j}^c)_\alpha^\dagger +  \kappa_b (\sigma)\, {y'}^{ij}_d (Q_{L,i} d_{R,j}^c)_\alpha +   \kappa_l (\sigma)\, {y'}^{ij}_l (L_i l_j^c)_\alpha^\dagger\right) \mbox{Tr} [P^\alpha \Sigma]  + h.c. \nonumber
\end{eqnarray}
where $D_\mu$ contains the EW gauge bosons, and we added the couplings of the Sp(4) singlet $\sigma$. The matrices $P^\alpha$ are spurions that project the pion matrix on its components transforming as a doublet of SU(2)$_L$.
As we shall see later, the $\sigma$ can also play the role of the Higgs boson, even though naively its mass is expected to be large.
The second line contains effective couplings of the condensate to the SM fermions. Such terms are necessary in order to give mass to the fermions in a similar way as Yukawa couplings do in the SM.
A possible origin of such terms can be traced back to four-Fermi interactions in the form (for the up-sector):
\begin{equation}
\mathcal{L}_{\rm EFCD} = - \frac{y_u^{ij}}{\Lambda_u^2}\, (Q Q)^\alpha (Q_{L,i} u_{R,j}^c)_\alpha^\dagger + h.c.
\end{equation}
As all the Yukawa terms have the same 4-Fermi origin, one may expect $\kappa_t = \kappa_b = \kappa_l$.

A detailed analysis of this Lagrangian can be found in~\cite{Cacciapaglia:2014uja}: here we will limit ourselves to listing the main results relevant for the phenomenology of the scalar sector.
First, the alignment of the vacuum generates masses for both the $W$ and $Z$, as well as  fermions:
\begin{equation}
m_W^2 = 2 g^2 f^2 \sin^2 \theta = \frac{g^2 v^2}{4}\,, \quad m_Z^2 = \frac{m_W^2}{\cos^2 \theta_W}\,, \qquad m_f = {y'}_f f \sin \theta = \frac{y'_f v}{2 \sqrt{2}}\,,
\end{equation}
where $v = 2 \sqrt{2} f \sin \theta$, and the relation between the $Z$ and $W$ masses is guaranteed by the custodial symmetry.
Additional small corrections arise from the singlet field $\sigma$ acquiring a vacuum expectation value, however such corrections will be neglected in the following.
Eq.~(\ref{eq:CCWZ}) also determines the couplings of the scalars, both pNGBs and $\sigma$, to the gauge bosons and SM fermions.
First it should be said that the first 3 pions are exact Goldstone bosons and are eaten by the massive $W$ and $Z$ (for $\theta \neq 0$), so they can be removed in the Unitary gauge. About the remaining two pions, they are both pNGBs and, while one of them behaves like a Higgs boson in the sense that it couples linearly to the massive states, the other is a singlet and only couples quadratically. We can therefore rename the two as $h$ and $\eta$:
\begin{equation}
\Sigma = e^{i Y^4 h/f + i Y^5 \eta/f} \cdot \Sigma_0\,.
\end{equation}
Expanding Eq.~(\ref{eq:CCWZ}) in the unitary gauge, one obtains
\begin{equation}
g_{hWW} = \sqrt{2} g^2 f \sin \theta \cos \theta = g_{hWW}^{\rm SM} \cos \theta\,, \qquad g_{hf\bar{f}} = \frac{y'_f}{\sqrt{2}} \cos \theta = g_{hf\bar{f}}^{\rm SM} \cos \theta\,,
\end{equation}
 while the couplings to the $Z$ are determined by custodial invariance.
Similarly, expanding
\begin{equation}
\kappa (\sigma) = 1 + \frac{\kappa^{(1)}}{4 \pi f} \sigma + \frac{1}{2} \frac{\kappa^{(2)}}{(4 \pi f)^2} \sigma^2 + \dots
\end{equation}
one finds the couplings of $\sigma$:
\begin{equation}
g_{\sigma WW} = \frac{\kappa^{(1)}_G}{4 \pi f} m_W^2 = g_{hWW}^{\rm SM} \tilde{\kappa}_G \sin \theta\,, \quad g_{\sigma f\bar{f}} = \frac{\kappa^{(1)}_f}{4 \pi f} m_f = g_{hf\bar{f}}^{\rm SM} \tilde{\kappa}_f \sin \theta\,, 
\end{equation}
where we have defined
\begin{equation}
\tilde{\kappa}_G = \frac{\kappa^{(1)}_G}{2 \sqrt{2} \pi}\,, \quad \tilde{\kappa}_f = \frac{\kappa^{(1)}_f}{\sqrt{2} \pi}\,, \quad \tilde{\kappa}_G^{(2)} = \frac{\kappa_G^{(2)}}{4 \pi^2}\,, \quad \quad \tilde{\kappa}_f^{(2)} = \frac{\kappa_f^{(2)}}{2 \pi^2}\,,
\end{equation}
for later convenience.
A summary of the couplings of the 3 scalars to the SM states normalised to the SM values can be found in Table~\ref{tab:couplings}.
It is also useful to complete the list with the couplings of two scalars to the fermions, which are absent in the SM but may be relevant for the pair production of the scalars at the LHC:
\begin{eqnarray}
g_{hhf\bar{f}} &=& -\frac{m_f}{v^2} \sin^2 \theta\,, \\
g_{\sigma \sigma f \bar{f}} &=& \frac{\kappa_f^{(2)}}{(4 \pi f)^2} m_f = \tilde{\kappa}_f^{(2)} \frac{m_f}{v^2} \sin^2 \theta\,, \\
g_{h \sigma f \bar{f}} &=& \frac{\kappa_t^{(1)}}{4 \pi f} \frac{m_f}{v} \cos\theta= \tilde{\kappa}_f \frac{m_f}{v^2} \sin \theta \cos \theta\,, \\
g_{\eta^2 f \bar{f}} &=& -\frac{m_f}{v^2} \sin^2 \theta\,.
\end{eqnarray}
The importance of such couplings for the Higgs pair production has been stressed in Ref.~\cite{Contino:2012xk}.

 \begin{table}[tb]
 \begin{center}
 \begin{tabular}{l|c|c|}
    & $WW$, $ZZ$ & $f\bar{f}$ \\
\hline
$h$ & $\cos\theta$ & $\cos\theta$ \\
$\sigma$ & $\tilde{\kappa}_G \sin\theta$ & $\tilde{\kappa}_f \sin\theta$ \\
$\eta$ & - & - \\
\hline
$hh$ & $\cos 2 \theta$ &   \\
$\sigma h$ & $\tilde{\kappa}_G \sin 2 \theta$ & \\
$\sigma \sigma$ & $\tilde{\kappa}^{(2)}_G \sin^2 \theta$ & \\
$\eta \eta$ & - $\sin \theta^2$ &  \\
\hline
\end{tabular} 
\end{center}
\caption{Coupling of one and two scalars to gauge bosons and fermions normalised to the SM value. The bilinear couplings to fermions are not reported here as they are absent in the SM.} \label{tab:couplings}
\end{table}

\subsection{The Higgs spectrum and fine-tuning}

The masses of the pNGBs are generated by operators that break explicitly the global symmetry both at tree and loop levels.
The potential used in Refs.~\cite{Cacciapaglia:2014uja,Galloway:2010bp} consists of 3 contributions:
\begin{equation}
V_{\rm scalars} = \kappa_G (\sigma)\, V_{\rm gauge} + \kappa_t^2 (\sigma)\, V_{\rm top} + \kappa_m(\sigma)\, V_m \,.
\end{equation}
The first two terms are generated by loops of the EW gauge bosons and the top, while the third comes from the mass term of the techni-quarks, which explicitly breaks SU(4)$\to$Sp(4).
Here we will summarise the main results. To further simply the analysis we neglect the gauge boson contribution. This is justified by the fact that it is  smaller than the top one.
Also, we omit the contribution of $\sigma$ to identify the symmetry breaking alignment with respect to the electroweak symmetry. 

First, we can compute the potential for $\theta$:
\begin{equation}
V(\theta) = {y'_t}^2 C_t\, \cos^2 \theta - 4 C_m\, \cos \theta + \mbox{constant}
\end{equation}
where $C_{t,m}$ are order 1 coefficients determined by the dynamics ($C_t$ is expected to be positive to match the sign of a fermion loop).
The minimum of the potential is given by
\begin{equation}
\cos \theta_{\rm min} = \frac{2 C_m}{{y'_t}^2 C_t}\,, \qquad \mbox{for}\;\; {y'_t}^2 C_t > 2 |C_m|\,.
\end{equation}
Note that a small $\theta$ can only be achieved for $2 C_m \to {y'_t}^2 C_t$: in order to reach the pNGB Higgs limit, one needs therefore to fine-tune two contributions in the potential which are of very different origins.
This is the only severe fine-tuning required in the model, if a small $\theta$ needs to be achieved.
Note also that in the limit of a small mass for the techni-fermions, $C_m \ll C_t$, the vacuum moves towards the TC limit $\theta = \pi/2$. Remarkably, a non fine-tuned (in $\theta$) realisation of a pNGB Higgs may occur if its nature is elementary \cite{Alanne:2014kea}, the reason being that the corrections to the potential, once the quadratic divergences are properly subtracted, derive from the dependence of the potential on the fourth power of the couplings (corresponding to logarithmically divergent corrections to the quartic coupling) rather than on the quadratic power (corresponding to the quadratically divergent contribution to the mass). It is also noteworthy that here we used an explicit mass term for the Techni-quarks to stabilise the potential, while in other models of composite (pNGB) Higgs in the market the stabilisation is due to quartic terms: as we just discussed, quartic terms are the dominant ones in elementary realisations of the pNGB Higgs, however they are subleading in composite realisations.

This potential also determines the masses of the pNGBs:
\begin{eqnarray}
m_{\chi_{1,2,3}}^2 &=& \frac{f^2}{4} \left(2 C_m  -  {y'_t}^2 C_t\cos \theta\right) \cos \theta = 0\,,\\
m_{h}^2 &=& \frac{f^2}{4} \left( 2 C_m\cos\theta -  {y'_t}^2 C_t \cos(2 \theta)\right) = \frac{{y'_t}^2 C_t f^2}{4} \sin^2 \theta\,,\\
m_{\eta}^2 &=&  \frac{f^2}{4} \left( 2 C_m \cos \theta + {y'_t}^2 C_t \sin^2 \theta \right)  =\frac{{y'_t}^2 C_t f^2}{4} \,,
\end{eqnarray}
where we have used the minimum condition to remove the dependence on $C_m$.
We notice here that, as expected, the new fundamental elementary fermion mass term gives the same mass (of order $f$) to all pions.
On the other hand, the top loop gives a mass of order $f$ to the pNGB Higgs, and a mass of order $f \sin \theta$ to the EW singlet.
This can be easily understood: the top couples via 4-fermi interactions to the techni-quarks doublet that transforms as a doublet of SU(2), thus the top loop will generate the usual divergent contribution to its mass that, following naive dimensional analysis, can be approximated as
\begin{equation} \label{eq:mh}
\Delta m_h^2 (\mbox{top}) = C \frac{{y'_t}^2}{16 \pi^2} \Lambda^2 = C {y'_t}^2 f^2\,.
\end{equation}
This large contribution, however, is cancelled by the contribution of the explicit mass at the minimum, so that the final value of the pNGB Higgs mass is
\begin{equation}
m_h^2 = \frac{{y'_t}^2 C_t f^2}{4} \sin^2 \theta = m_\eta^2 \sin^2 \theta = \frac{C_t m_{\rm top}^2}{4}\,.
\end{equation}
Note that it would be enough to have $C_t \sim 2$ to generate the correct value for the Higgs mass. The value of $C_t$ is not a free parameter, but it can be determined by the dynamics.
At present, no calculation of such coefficients is available.
Nevertheless, no additional fine-tuning is in principle necessary for the Higgs mass, once the fine-tuning in the alignment is paid off.
The relation between the masses of $h$ and $\eta$ also survives after the gauge corrections are included, however it can easily be spoiled by other corrections, like for instance the mixing between $h$ and $\sigma$.
Finally, the pions eaten by the $W$ and $Z$ are massless on the correct vacuum, as expected for exact Goldstone bosons.

The parametric smallness of the pNGB Higgs mass can also be understood in terms of symmetries. The 3 Goldstone bosons eaten by $W$ and $Z$ 
are always massless, for any value of $\theta$. Therefore, if we go continuously to the limit $\theta\to 0$, where the EW symmetry is restored, the 
mass of the pNGB Higgs must also vanish in order to reconstruct a complete massless SU(2) doublet.
The same argument cannot be applied to $\eta$, which is a singlet unrelated to EW symmetry breaking.

In the natural presence of the singlet $\sigma$ the spectrum is affected. In fact, $\sigma$ mixes with $h$ (but not with $\eta$), as they share the same quantum numbers. A detailed description of the mass matrix can be found in~\cite{Cacciapaglia:2014uja}.
Here, we will keep the discussion general, so we will simply replace $h$ and $\sigma$ by the mass eigenstates $h_{1,2}$, where the lighter states $h_1$ is identified with the observed Higgs at $m_{h_1} = 125$ GeV:
\begin{equation}
\begin{pmatrix}
  h_1 \\
  h_2 \\
  \end{pmatrix}
  =
 \begin{pmatrix}
  c_{\alpha} & s_{\alpha} \\
  -s_{\alpha} & c_{\alpha} \\
  \end{pmatrix}
  \begin{pmatrix}
  h \\
  \sigma \\
  \end{pmatrix}\,.
  \label{eq:mixing h sigma}
\end{equation}
Both the mass $m_{h_2}$ and the mixing angle $\alpha$ will be considered here as independent free parameters.
It should only be reminded that $\alpha \to 0$ for $\theta \to 0$, as the EW symmetry is not broken in that limit, and also $\alpha\to \pi/2$ for $\theta \to \pi/2$ as a global U(1) subgroup will prevent mixing in the TC limit~\cite{Ryttov:2008xe} (and we need to associate the observed Higgs with $\sigma$).
The sign of $\alpha$ is not determined, however the analysis in \cite{Cacciapaglia:2014uja} shows that the mass of the light state will generically receive a negative correction from the mixing, that is reduced with respect to the prediction in Eq.~(\ref{eq:mh}).
We can therefore consider that
\begin{equation}
m_\eta > \frac{m_{h_1}}{\sin \theta}\,.
\end{equation}
In the phenomenological results of Section~\ref{sec:eta}, we will assume the equality as a limiting scenario.

\subsection{Trilinear scalar self-interactions}
We present here the trilinear couplings among scalars, which are relevant for the pair production of the discovered Higgs~\footnote{The couplings are proportional to the pNGB mass $m_h$ and not to the physical Higgs mass $m_{h_1}$. Thus these couplings should be compared to the SM value $g_{h^3}^{SM} = 3 m_{h_1}^2/v$.}  
\begin{eqnarray}
g_{h^3} &=& \frac{3 m_{h}^2}{v}  \cos \theta\,, \\
g_{\sigma h^2} &=& \frac{m_{h}^2}{v} \frac{1}{\sin \theta} \left( \tilde{\kappa}_m^{(1)} \cos^2\theta - 2 \tilde{\kappa}_t \cos (2 \theta) \right)\,, \\
g_{\sigma^2 h} &=& \frac{m_{h}^2}{v} \frac{2 \cos \theta}{\sin \theta} \left( \tilde{\kappa}_m^{(2)} - (\tilde{\kappa}_t^{(2)} + \tilde{\kappa}_t^2) \right)\,,
\end{eqnarray}
and of $\eta$:
\begin{eqnarray}
g_{h\eta^2} &=& \frac{m_{h}^2}{v}\cos\theta\,,\\
g_{\sigma \eta^2} &=& \frac{m_{h}^2}{v} \frac{1}{\sin \theta} \left( \tilde{\kappa}_m^{(1)} \cos^2\theta + 2 \tilde{\kappa}_t \sin^2 \theta \right)\,,
\end{eqnarray}
where we defined, for convenience,
\begin{equation}
\tilde{\kappa}_{m}^{(1)} = \frac{\kappa_{m}^{(1)}}{\sqrt{2} \pi}\,, \quad \tilde{\kappa}_{m}^{(2)} = \frac{\kappa_{m}^{(2)}}{2 \pi^2}\ , 
\end{equation}
and we are working in the non-diagonalised scalar basis.

It is interesting to notice that the couplings of $\sigma$ diverge for small $\theta$: this is a sign that they are proportional to the condensation scale $f$ and thus increase for increasing condensation scale. The trilinear coupling of $\sigma$ cannot be determined as it comes directly from the strong dynamics. It should therefore be considered as an additional free parameter, also proportional to the condensation scale $f$.

\subsection{Bounds from EWPTs}
\label{sec:EWPT}
The precise determination of the oblique corrections is a delicate issue in composite extensions of the SM. A well defined procedure must be employed that allows to clearly disentangle the intrinsic contribution stemming from strong dynamics from the one coming from the genuine SM contribution \cite{Foadi:2012ga}.   Once such a procedure is established, an estimate from the strongly coupled sector is needed. First principle lattice simulations are the primary method to determine this contribution. However, as a very rough estimate, one can use the one-loop contribution from the fundamental fermions with heavy constituent mass terms.  In this case for one SU(2)$_L$ doublet we have $\Delta S = 1/(6\pi)$. In the fundamental model under consideration, this translates into the following contribution
\begin{equation} \label{eq:naiveS}
\Delta S_{UV} = \frac{\sin^2 \theta}{6\pi}
\end{equation}
for each fundamental doublet \footnote{We note that this estimate is modified when the underlying dynamics is near conformal  because of the violation of the second Weinberg's sum rule \cite{Appelquist:1998xf}. There is a limit, however, when this estimate turns into a precise result. This occurs close to the upper limit of the conformal window \cite{Sannino:2010ca}  provided the correct kinematical limits are chosen. The two-loop contributions have been computed in \cite{DiChiara:2010xb} where it is clearly shown how the S parameter increases when moving deeper into the nonperturbative region.  In the non-perturbative regimes recent comprehensive holographic estimates have appeared \cite{Jarvinen:2015ofa}. }.
The reason for the presence of the $\sin^2 \theta$ term can be understood in terms of symmetries: in the radial composite Higgs limit $\theta \to \pi/2$, the fundamental fermions pick up a dynamical mass from the condensate which is aligned with the EW breaking direction, thus the calculation satisfies some of the assumptions in~\cite{Peskin:1991sw}; on the other hand, in the limit $\theta\to 0$, the EW symmetry is recovered and the $S$ parameter must vanish.
The power is understood in terms of masses: in fact, it is expected to be proportional to the square of the ratio of the dynamical mass aligned to the EW breaking direction, $\sim f \sin \theta$, and the total dynamical mass of the fermions, $\sim f$.
This expectation is also confirmed by an operator analysis of this contribution, as shown in~\cite{Galloway:2010bp}.
The strongly interacting contribution to the $T$ parameter vanishes because the dynamics respects the $SU(2)_V$ custodial symmetry.

The underlying strong dynamics contribution must then be matched with the  important one coming from the quantum corrections in the effective Lagrangian for the lightest states considered here. 
We will use a more naive way to estimate the total correction: we explicitly include the contribution of the loops of the lightest composite states~\footnote{The $\eta$ does not contribute: in fact, its couplings can only generate corrections to the masses and, because of the custodial symmetry, such corrections do not enter the $T$ parameter.}, i.e. the 125 GeV Higgs $h_1$, which contributes due to the modified couplings to gauge bosons, and the heavier ``Higgs'' $h_2$.
Then, we will assume that the contribution of the heavier resonances can be approximated by the Techni-quark loop in Eq.~(\ref{eq:naiveS}), as one would expect if the contribution were dominated by the lightest vector and axial resonances. 
The net effect can be estimated starting from the contribution of the Higgs loops and summarise the results in the following:
\begin{eqnarray}
\Delta S &=& \frac{1}{6 \pi} \left[ (1-k_{h_1}^2) \ln \frac{\Lambda}{m_{h_1}} - k_{h_2}^2 \ln \frac{\Lambda}{m_{h_2}} + N_D \sin^2 \theta \right]\,, \\
\Delta T &=& - \frac{3}{8 \pi \cos^2 \theta_W} \left[ (1-k_{h_1}^2) \ln \frac{\Lambda}{m_{h_1}} - k_{h_2}^2 \ln \frac{\Lambda}{m_{h_2}} \right]\,,
\end{eqnarray}
where 
\begin{equation}
k_{h_1} = \cos (\theta-\alpha) + (\tilde{\kappa}_G -1) \sin \theta \sin \alpha\,, \quad k_{h_2} = \sin (\theta-\alpha) +(\tilde{\kappa}_G -1) \sin \theta \cos \alpha\,, 
\end{equation}
and $N_D$ is the number of techni-fermion doublets ($N_D = 2$ for SU(2)$_{\rm FCD}$, and $2N$ for Sp($2N$)$_{\rm FCD}$). In this analysis we assumed the presence of physical cutoff $\Lambda$ to be identified with the next massive state.  The dependence on the cutoff emerges because the scalar loop contributions are divergent, as a sign of the effective nature of the Lagrangian. 
The divergence is corrected once the proper matching to the underlying UV dynamics is taken into account. In our phenomenological estimates, we will use
\begin{equation}
\Lambda  = 4 \pi f = \frac{\sqrt{2} \pi v}{\sin \theta}\,,
\end{equation}
which is very close to the mass of the spin-1 resonances as shown by first-principle lattice simulations \cite{Lewis:2011zb,Hietanen:2013fya,Hietanen:2014xca,Arthur:2014lma}. We also added to $\Delta S$ the naive strongly coupled contribution that should partially take into account the heavier states. 
This estimate is clearly naive but should capture at least the correct order of magnitude of the corrections. 
It should be stressed that a more appropriate calculation should be employed if one wanted to use Lattice calculations of the contribution of the strong dynamics to $S$, as thoroughly discussed in \cite{Foadi:2012ga}, where one finds also the discussion of the needed counterterms in the effective Lagrangian.

\section{Constraints from the Higgs coupling measurements and EWPTs} \label{sec:ewpt}

Even though the couplings of the Higgs boson have been measured with a precision that is at the level of 10\% in the best cases, the fact that they are close to the SM values poses significant constraints on the scenarios of composite Higgs.
In this work, we use the final analyses of the data collected in 2011 and 2012 by the LHC collaborations, CMS \cite{Khachatryan:2014jba} and ATLAS \cite{Aad:2014xzb,Aad:2014eva,ATLAS:2014aga,Aad:2015vsa,Aad:2014eha}, to extract the constraints from the Higgs couplings. The results of the experimental analyses are provided as exclusion contours in terms of signal strengths, and treated in the way described in \cite{Cacciapaglia:2012wb}. These experimental plots represent regions allowed at $68\%$ confidence level (C.L.) by the analyses, in the plane of cross sections rescaling factors for the  main Higgs decay channels $H \to \gamma \gamma, WW^*, Z Z^* , \bar \tau \tau, \bar b b$, under the assumption that $W$ and $Z$-strahlung (VH) and vector boson fusion (VBF) modes are rescaled by the same factor, as well as  the gluon fusion and $t\bar{t}H$.  We fitted these lines as ellipses, therefore extrapolating the $\chi^2$ for each channel as a paraboloid, i.e. approximating the likelihood functions with a gaussian. The exception to this procedure is that in the ATLAS measurement,  $H \to \bar b b$  is only selected  via the VH channel, thus we included its one dimensional signal strength and uncertainty into the likelihood function.  Through the $\chi^2$ function  we will  determine the best fit point  and   we can then use the reconstructed  quantity   $\Delta \chi^2  = \chi^2 - \chi^2_{min}$  to draw the exclusion limits. This method has been validated to reproduce the experimental results~\cite{Flament}.

\subsection{The composite pNGB Higgs limit}

Here we assume $\sigma$ to decouple and we identify the discovered Higgs with the pNGB $h$. This limit corresponds to the case $\alpha = 0$ and $\tilde{\kappa}_G = \tilde{\kappa}_t = 0$.
In this limit both the Higgs couplings and EWPTs depend only on  $\theta$, thus allowing us to extract an upper bound on the value of this angle. The limits at 3$\sigma$ are summarised in the following table:
\begin{center}
\begin{tabular}{l|c|c|c|c|}
  & Higgs couplings & EWPTs - SU(2)$_{\rm FCD}$ & EWPTs - Sp(4)$_{\rm FCD}$ & EWPTs - Sp(6)$_{\rm FCD}$ \\ 
\hline
$\theta <$ &   $\begin{array}{c} 0.71\;\; (\mbox{CMS}) \\ 0.61\;\;(\mbox{ATLAS}) \end{array}$ & $0.239$ & $0.227$ & $0.216$\\
\hline
\end{tabular}
\end{center}
The numbers show that the bound from EWPTs is much stronger than the bounds from the Higgs couplings, and points to values $\sin \theta \leq 0.2$. This value is consistent with bounds obtained in other models of pNGB Higgs~\cite{Barbieri:2007bh}.
There is also a mild dependence on the number of doublets in the dynamical model, thus signaling that the bound is dominated by the contribution of the Higgs boson.

\begin{figure}[tb!]
\center
\includegraphics[scale=0.95]{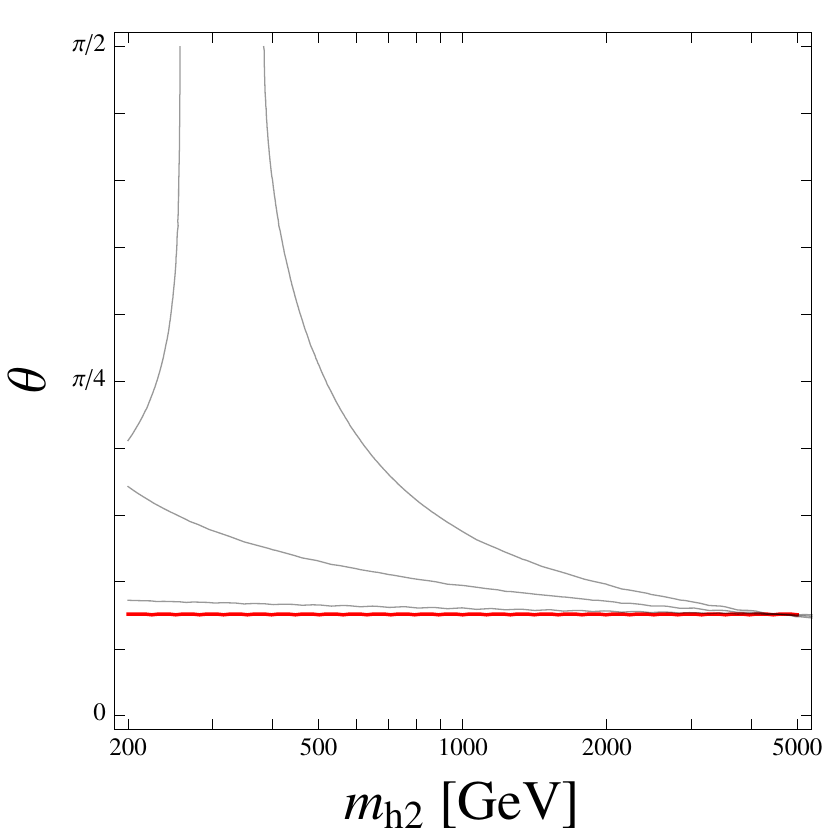}
\caption{Upper bound on $\theta$ as a function of the mass of $\sigma$. The red curve corresponds to the decoupling limit $\theta < 0.239$, while for the other lines correspond to $\tilde{\kappa}_G = 0.5$, $1$ and $1.2$, while we keep $\alpha = 0$ and $N_D = 2$.}
\label{fig:EWPTpNGB}
\end{figure}

In this analysis we ignored the presence of $\sigma$.  However, one can imagine a situation where $\alpha \sim 0$ with a large value of $\tilde{\kappa}_G$. In this limit the mass eigenstate $h_2$ can therefore affect EWPTs. This situation can be achieved because the mixing between $\sigma$ and $h$ is mostly generated via the coupling of $\sigma$ to the top and the mass term in the potential for the pNGBs, while the bounds are only sensitive to the coupling to gauge bosons.
Even when the mixing vanishes  the $\sigma$ state still affects the EWPTs as shown in Figure~\ref{fig:EWPTpNGB}.  Here we plot the upper bound on $\theta$ as a function of the mass of the heavier scalar mass for various values of $\tilde{\kappa}_G$.
One can see that a non-zero value of the couplings can relax the bound and, for $\tilde{\kappa}_G > 1$ there is a range in mass where the EWPTs cannot bound $\theta$. Additional constraints arise form the measured couplings of the discovered Higgs and from direct searches on the heavier $h_2$ (the latter will be discussed in Section~\ref{sec:h2}).
This window is interesting because it shows how to relax the EWPTs  but will not be pursued here because it  requires a certain tuning of the $\sigma$  couplings to obtain $\alpha \sim 0$.

\subsection{The Technicolor limit}

\begin{figure}[tb!]
\center
\includegraphics[scale=0.8]{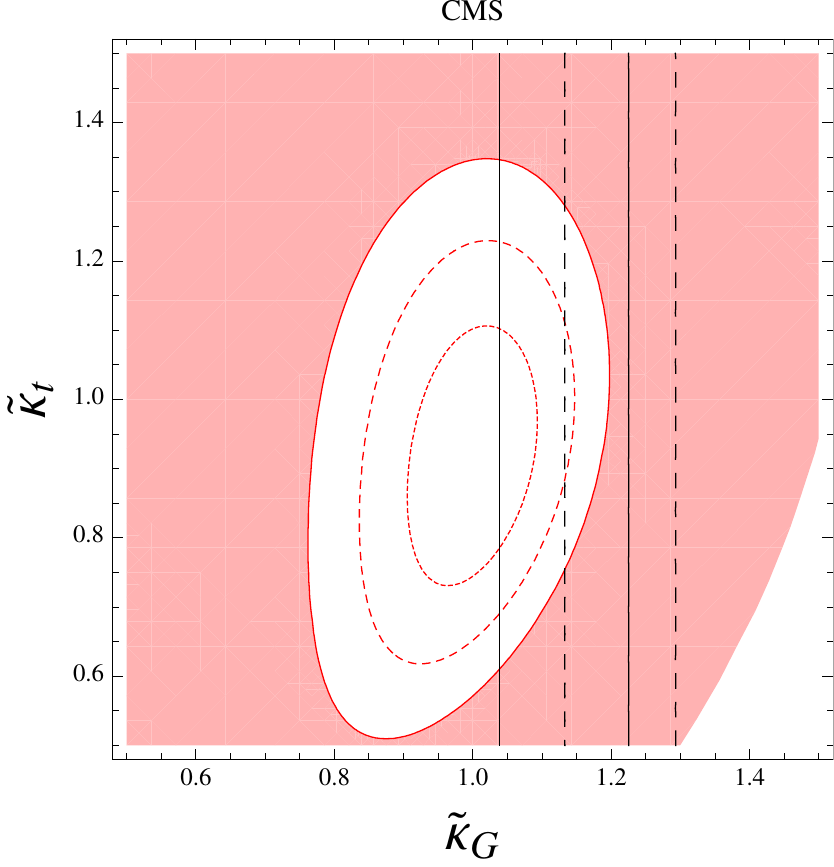}
\includegraphics[scale=0.8]{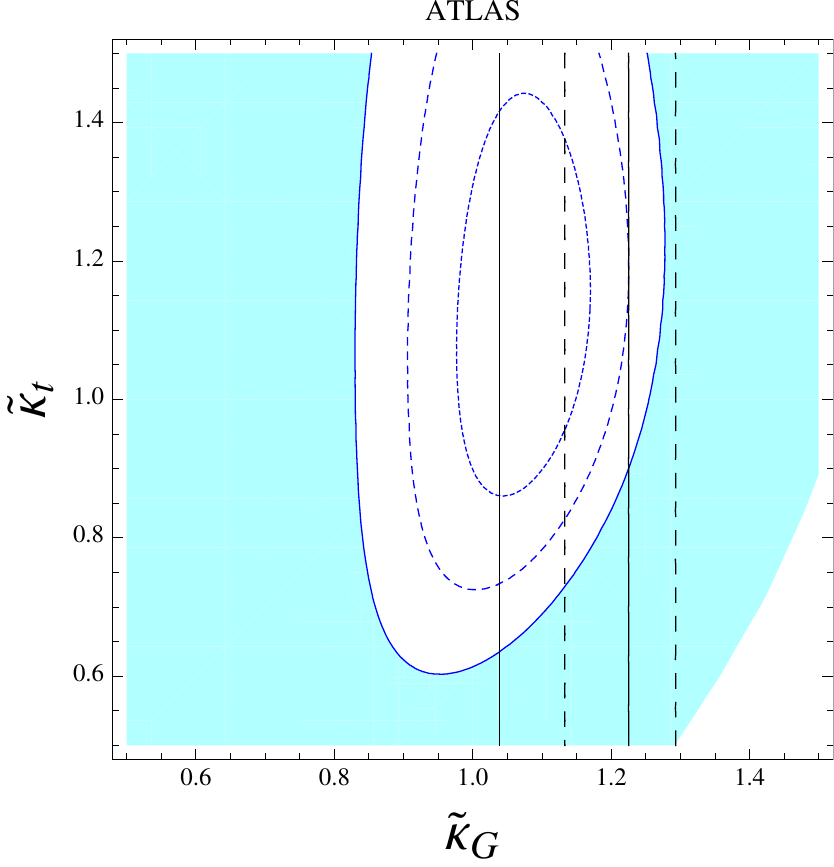}
\caption{Region allowed by the Higgs couplings in the TC limit $\theta = \alpha = \pi/2$, with 1, 2 and 3$\sigma$ contours (left panel CMS bound and right panel ATLAS). The regions within the vertical lines are allowed by EWPTs at 3$\sigma$ for SU(2)$_{\rm FCD}$ (solid) and Sp(4)$_{\rm FCD}$ (dashed).}
\label{fig:EWPTTC}
\end{figure}

Another interesting limit occurs for $\theta = \pi/2$, i.e. the TC limit. In this case, the mixing vanishes and the Higgs is associated with 
$\sigma$, i.e. $\alpha = \pi/2$.
The pNGB $h$ decouples and, together with $\eta$, may play the role of dark matter~\cite{Ryttov:2008xe}, while the couplings of the 125 GeV Higgs depend on the details of the underlying dynamics and are associated to the $\tilde{\kappa}$ parameters.
The correct value of the Higgs mass can be achieved via a cancelation between the dynamical mass, of the order of a TeV, and loop contributions from explicit breaking of the global symmetry~\footnote{A lighter mass can be achieved by considering underlying dynamics that is not QCD-like \cite{Hong:2004td,Sannino:2004qp}, or a near-conformal one \cite{ Dietrich:2005jn,Dietrich:2006cm}. This situation can be achieved in the model under consideration by adding a small number of fermions in the adjoint representation \cite{Ryttov:2008xe} of SU(2)$_{\rm FCD}$. }, such as the top loops~\cite{Foadi:2012bb}.
Assuming the mass gets the correct value, we can compute the bounds on the couplings of $\sigma$ to gauge bosons $\tilde{\kappa}_G$ and fermions $\tilde{\kappa}_t$ (we are explicitly assuming that all fermions couple in the same way, i.e. $\tilde{\kappa}_t = \tilde{\kappa}_b = \tilde{\kappa}_l$).
The results are shown in Fig.~\ref{fig:EWPTTC}, where we show 1, 2 and 3$\sigma$ contours from the measured Higgs couplings from CMS (left panel) and ATLAS (right panel).
A fairly large region around the SM limit $\tilde{\kappa}_G = \tilde{\kappa}_t = 1$ is still open.
In principle, there is no reason for these couplings to be close to the ones of the SM Higgs, however it is fascinating that this happens for the $\sigma$ meson in QCD~\cite{Belyaev:2013ida}. 
We also compare this allowed region with the bound from EWPTs, which is only dependent on $\tilde{\kappa}_G$. The vertical lines delimit the allowed region for two choices of $N_D$, corresponding to SU(2)$_{\rm FCD}$ and Sp(4)$_{\rm FCD}$. For SU(2)$_{\rm FCD}$, which corresponds to two doublets, a substantial overlap exists, pointing to larger couplings to gauge bosons and smaller couplings to fermions with respect to the SM values. This effects should become measurable once more precise data on the Higgs couplings are available. The intersection becomes smaller for Sp(4)$_{\rm FCD}$, which has 4 doublets, while larger Sp($N$)$_{\rm FCD}$ are clearly disfavoured as EWPTs push the parameters in a region excluded by the Higgs coupling measurements.
Our results clearly show that the TC limit is still allowed, provided that the correct value of the mass can be achieved.

\subsection{General case}

\begin{figure}[tb!]
\center
\includegraphics[scale=0.8]{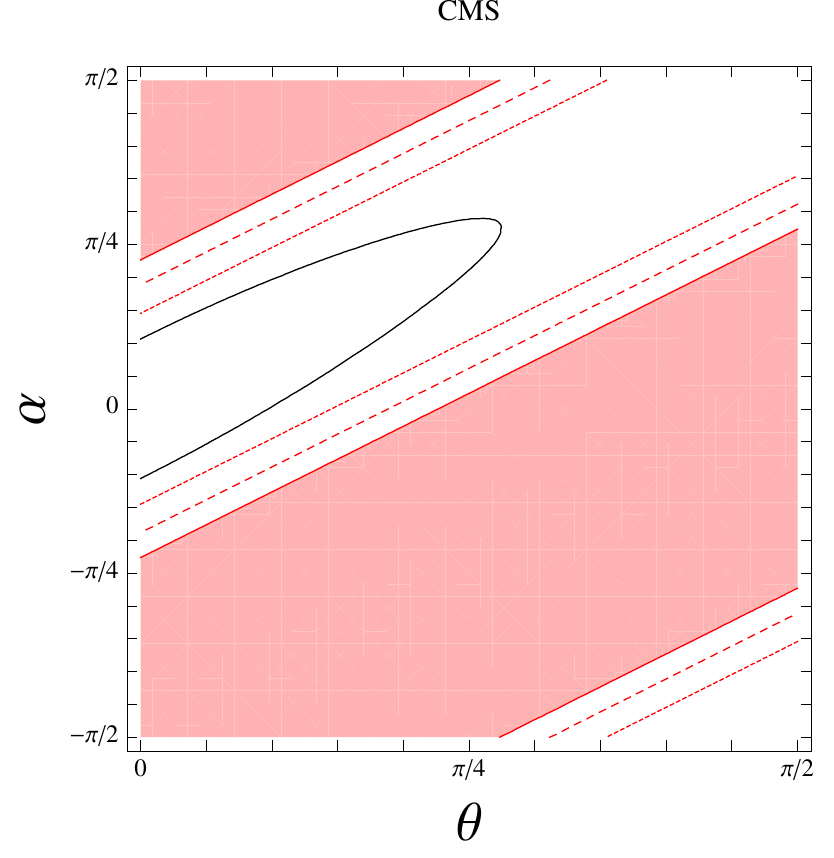}
\includegraphics[scale=0.8]{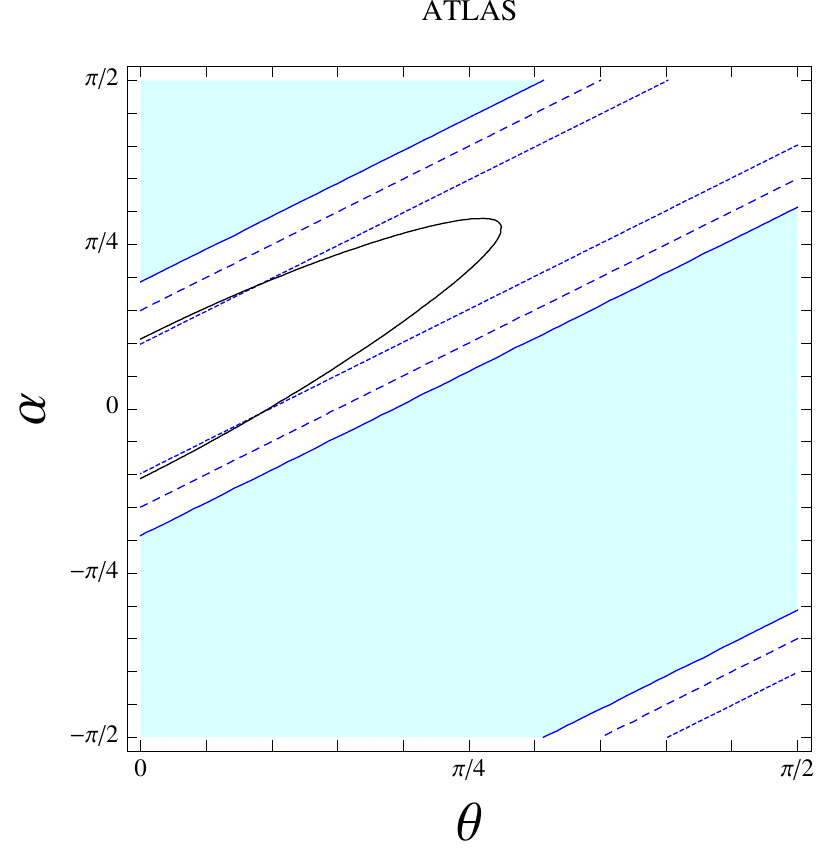}
\caption{Region allowed by the Higgs couplings for $\tilde{\kappa} = 1$ and $m_{h_2}=1$ TeV (left panel CMS bound and right panel ATLAS). The 
black line indicates the 3$\sigma$ bound from EWPTs in the SU(2)$_{\rm FCD}$.}
\label{fig:EWPTk10}
\end{figure}

\begin{figure}[tb!]
\center
\includegraphics[scale=0.8]{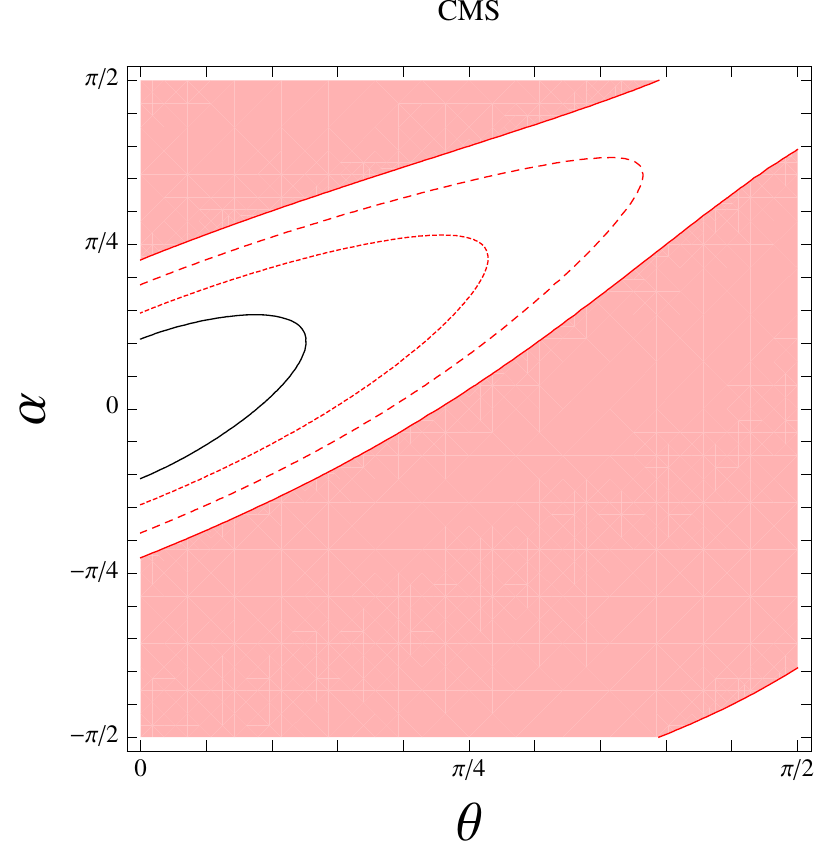} \includegraphics[scale=0.8]{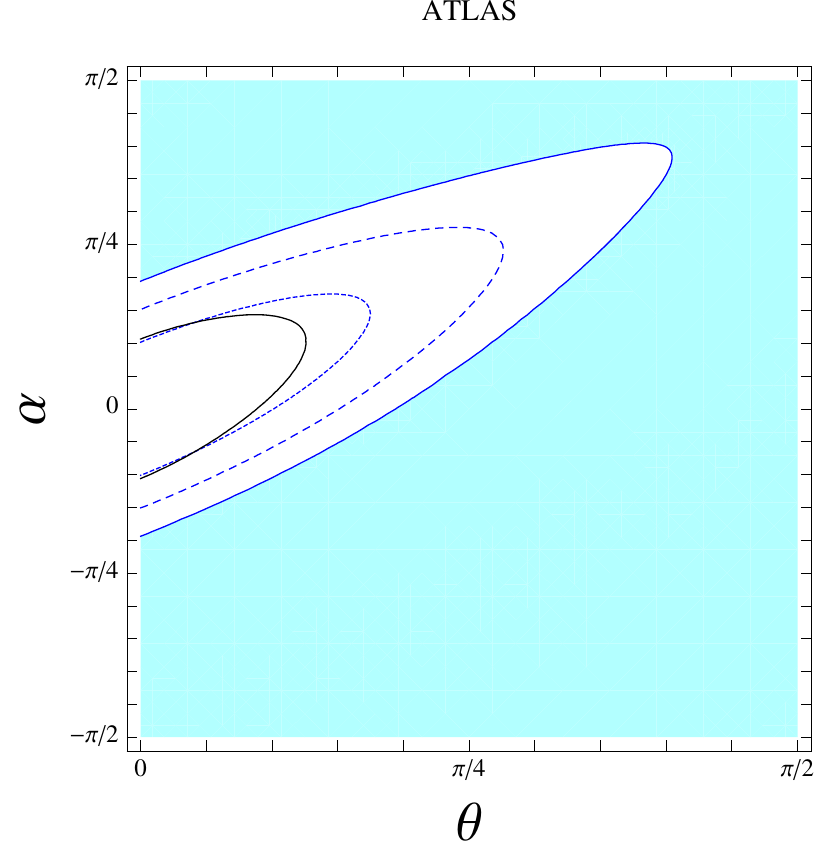}
\includegraphics[scale=0.8]{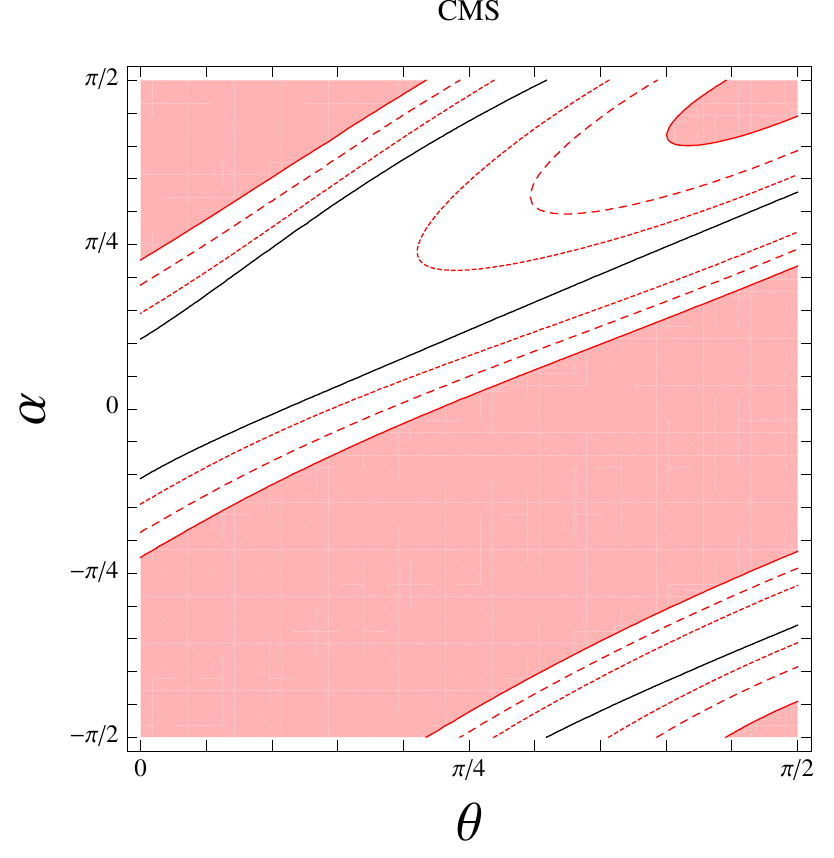} \includegraphics[scale=0.8]{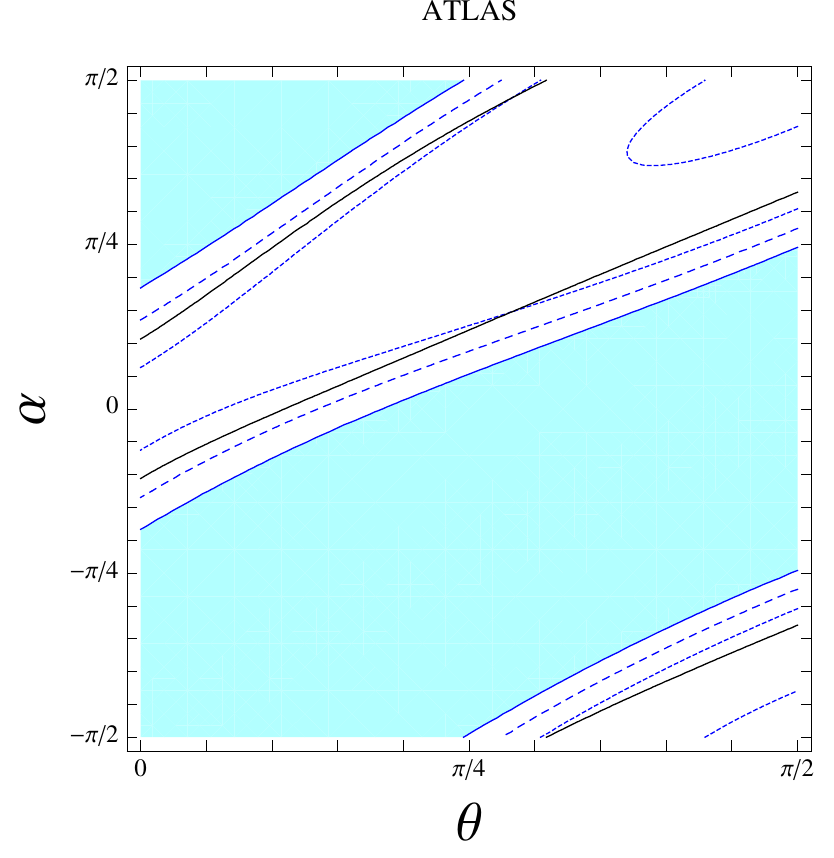}
\caption{Region allowed by the Higgs couplings for $m_{h_2}=1$ TeV (left panels CMS bound and right panels ATLAS), with 
$\tilde{\kappa} = 0.8$ (top row) and $\tilde{\kappa} = 1.2$ (bottom row). The black line indicates the 3$\sigma$ bound from EWPTs for 
SU(2)$_{\rm FCD}$.}
\label{fig:EWPTk}
\end{figure}

We now turn our attention to the general case. To reduce the number of unknown parameters, we fix the $\sigma$ couplings as follows: $\tilde{\kappa}_G = \tilde{\kappa}_t = \tilde{\kappa}$. We also fix the mass of the heavier Higgs $h_2$ and plot the bounds in the plane $\theta$--$\alpha$.
In Fig.~\ref{fig:EWPTk10}, we show the bounds in the case $\tilde{\kappa} = 1$, and $m_{h_2} = 1$ TeV.
The plot shows a degeneracy in the bounds from the Higgs couplings due to the fact that the couplings of both the light and heavy Higgses depend only on the difference $(\theta - \alpha)$.
On the other hand, EWPTs, in absence of any near-conformal dynamics \cite{Appelquist:1998xf} or sources of isospin breaking, as it is well known do prefer small $\theta$ cutting out the TC corner.
Interestingly, however, we observe a novel way to loosen the bound on $\theta$, by allowing, for example for a positive mixing angle $\alpha$ allowing for values of $\theta$ up to $\pi/4$. This is an interesting result since it would reduce the level of fine-tuning to achieve in these models either a pure pNGB or TC limit.

The situation is qualitatively different for values of $\tilde{\kappa}$ different from 1, as shown in Fig.~\ref{fig:EWPTk}.
For couplings smaller than unity, the allowed regions shrink, as the contribution of the heavy Higgs, which tends to compensate for the modification of the Higgs couplings, becomes less important.
For larger couplings $\tilde{\kappa} > 1$, the situation is very different: the EWPT allowed regions expand until the TC limit is reached, while the Higgs coupling bounds tend to shrink. In the bottom row of Fig.~\ref{fig:EWPTk}, drawn for $\tilde{\kappa} = 1.2$, we see that the TC limit is at odds with the (CMS) measurements of the Higgs couplings  but not with EWPTs. This is due, however, to our choice of $\tilde{\kappa}_G = \tilde{\kappa}_t$. We have shown in the previous section that smaller values of $\tilde{\kappa}_t$ would reconcile the Higgs couplings with the experimental measurements.

\section{Phenomenology of the singlet $\eta$} \label{sec:eta}

The pNGB $\eta$ is a light new state that appears in this model: the masses of spin-1 states have been computed on the lattice giving values above $2\div 3$ TeV \cite{Lewis:2011zb,Hietanen:2013fya,Hietanen:2014xca,Arthur:2014zda} since they grow inversely proportional to $\sin \theta$ \cite{Cacciapaglia:2014uja} . It is therefore interesting to assess the prospects for the discovery of this state at the LHC.
Note also that we studied the minimal case where, together with the Higgs(es), only one new pNGB appears: in non-minimal cases the $\eta$ will be accompanied by additional scalars, some of which may carry charge and EW quantum numbers.

This state has particular properties compared to the other pNGBs (i.e. the Higgs and the Goldstones eaten by the $W^\pm$ and $Z$).
The effective Lagrangian for the pNGBs in Eq.~(\ref{eq:CCWZ}) does not contain any coupling with an odd number of $\eta$ fields, thus showing an apparent symmetry
\begin{equation}
	\eta \to -\eta
	\label{eq:etaparity}
\end{equation}
that would prevent the $\eta$ from decaying. 
At the level of the Goldstone matrix $\Sigma$, this symmetry transformation can be expressed as
\beq \Omega \cdot \Sigma (h, - \eta) \cdot \Omega^T
= \Sigma (h, \eta)\,, \quad  \Omega = \left(
{\begin{array}{cc}
   {} & \sigma_2  \\
   \sigma_2 & {}  \\
\end{array}} \right)\,,
\eeq
where $\Omega$ is a transformation belonging to the unbroken Sp(4).
The action of this transformation on the gauged generators~\footnote{In fact, the off-diagonal blocks can contain any linear combination of the 3 Pauli matrices: we chose $\sigma_2$ because it allows for simpler transformation properties on the gauged generators of SU(4).} is
\beq \Omega^\dagger \cdot S^i_L \cdot \Omega =S^i_R \,,\eeq
which corresponds to exchanging the generators of $SU(2)_L$ with the
generators in $SU(2)_R$. About the top Yukawa, the coupling can be
written as:
 \beq ( t_L, b_L,
0, 0) \cdot \Sigma \cdot (0,0, t_R,0)^T \eeq
and the transformation rules are:
\beq (t_L, b_L, 0, 0)\cdot \Omega &=& (0,0, -i b_L, i t_L)\,,
\\
\Omega^T \cdot (0,0,t_R,0) &=& (0,i t_R, 0,0)\,,
 \eeq
where the SU(2)$_L$ doublet is transformed into an SU(2)$_R$
anti-doublet, and same for the incomplete SU(2)$_R$ doublet containing $t_R$.
This exchange is compatible with the exchange between the SU(2)$_L$ and SU(2)$_R$ generators, seen above.
The elementary fermions also pick up a complex phase, which is not physically relevant.

The parity changing sign to $\eta$ can, therefore, be though of as a systematic exchange of the two SU(2)'s, and this does not change the physical couplings of the pNGB Higgs $h$ nor of $\eta$, at the level of the leading order Lagrangian. The reason for this is that $\eta$ is a singlet, while $h$ couples to the symmetry breaking which is invariant under the exchange (being custodial invariant).

The action of $\Omega$ on the mass term $M_Q$ in Eq.~\eqref{eq:MQ} reads: 
\beq
\Omega^T \cdot M_Q \cdot \Omega = \Omega^T \cdot \left( \begin{array}{cc}
\mu_L i \sigma_2 & 0 \\
0 & \mu_R i \sigma_2
\end{array} \right) \cdot \Omega = - \left( \begin{array}{cc}
\mu_R i \sigma_2 & 0 \\
0 & \mu_L i \sigma_2
\end{array} \right)\,.
\eeq
Once again, the two masses corresponding to the SU(2)$_L$ and SU(2)$_R$ doublets are exchanged however with an additional minus sign, i.e. $\mu_{L/R} \to - \mu_{R/L}$.
The mass can be split into two terms:
\beq
M_Q = \frac{\mu_L - \mu_R}{2} \left( \begin{array}{cc}
i \sigma_2 & 0 \\
0 & - i \sigma_2 \end{array} \right) + \frac{\mu_L + \mu_R}{2} \left( \begin{array}{cc}
i \sigma_2 & 0 \\
0 & + i \sigma_2 \end{array} \right)\,.
\eeq
The first term, proportional to $\Sigma_B$, is even under the exchange, while the second is odd.
From this analysis we can deduce that the only spurion that breaks the $\eta$ parity explicitly is the second piece of the mass term, thus there will be breaking terms proportional to $\mu_L + \mu_R$.
Note that, for the analysis to be consistent, one would expect $\mu_L + \mu_R \ll \mu_L - \mu_R$, else the vacuum would align in a different direction.
Additional operators containing linear couplings of the $\eta$ will be generated at higher order, as we will show below.

\subsection{Linear couplings to fermions}

At leading order, there exists a unique operator which contains a linear
$\eta$-$f$-$f$ coupling~\footnote{In models with top partners~\cite{Barnard:2013zea}, linear couplings can also be generated via couplings to composite fermions~\cite{Gripaios:2009pe}.},
 generated by the mass and the top Yukawa~\cite{Galloway:2010bp}:
\beq
 \mathcal{O}_1  &=& \left( {Qt^c } \right)_\alpha ^\dag  {\rm Tr} \left[ {M_Q \Sigma P^\alpha  \Sigma } \right] 
 \eeq
where $\alpha, \beta$ are SU(2)$_L$ indices, and $M_Q$ is the mass matrix for the techni-quarks in Eq. (\ref{eq:MQ}).
Expanding, we obtain:
\beq  \mathcal{O}_1 &=& \left( {\mu_L -
\mu_R } \right)\cos \theta \sin \theta ~ t_L t_R^c  \nonumber \\
&+& \frac{1}{{2\sqrt 2 f}}\left[ {h\left( {\mu_L  - \mu_R }
\right)\cos 2\theta + i\eta \left( {\mu_L  + \mu_R } \right)\sin
\theta } \right] t_L t_R^c + \cdots \eeq
This operator generates a correction to the mass of the top (and coupling of the Higgs), together with a linear coupling of $\eta$. Notice however that they are proportional to different combinations of the masses: in particular, in the limit where the mass respects Sp(4), i.e. $\mu_R = - \mu_L$, the coupling to $\eta$ vanishes.
This coupling is therefore unrelated to the physics in the Higgs potential, and it is expected to be small, i.e. $\mu_L + \mu_R \ll \mu_L - \mu_R$, else the symmetry breaking patters SU(4)/Sp(4) that gives rise to a pNGB Higgs would be distorted. In the following we will estimate the maximum size of this operator in a different way from~\cite{Galloway:2010bp}. In fact, the operator gives a contribution to the top mass of the order
\beq
\delta m_t = C_1 \left( {\mu_L -
\mu_R } \right)\cos \theta \sin \theta
\eeq
where $C_1$ is an order  unity coefficient. This contribution can, in principle, be as large as the top mass. However, if that were the case, the contribution of the top coupling to the scalar potential would be affected and therefore it would modify the analysis performed so far.  
To be on the conservative side, we require that the corrections stemming from the operator above are at most 10\% of the top mass. We will consider a stronger effect in a future work.  The coupling of the $\eta$ can therefore be written as:
\begin{equation} \label{eq:getaff}
 g_{\eta t \bar{t}} = C_1 \frac{\mu_L + \mu_R}{2 \sqrt{2} f} \sin \theta \sim \tau \frac{10\% m_t}{v} \frac{\sin \theta}{\cos \theta}\sim 0.01 \frac{m_t}{v} \tan \theta \,,
\end{equation}
where $\tau = \frac{\mu_L + \mu_R}{\mu_L - \mu_R}$ parametrises the explicit breaking of Sp(4), and we have assigned to it a maximum value of $0.1$.
A similar operator can be written for all SM fermions, so that one can expect a similar coupling with, of course, the top mass replaced by the mass of the specific fermion considered. We also expect a similar overall coefficient. 
At higher order in the SM gauge couplings, or more generally in the chiral expansion, we expect several new operators. To elucidate this point we show two new operators containing a linear coupling of $\eta$ to fermions:
\beq
\mathcal{O}_2 &=& \left( {Qt^c } \right)_\alpha ^\dag\; \sum_i g_i^2 {\rm Tr} [S_i \Sigma S_i^\ast M_Q  \Sigma P^\alpha]\,, \\
\mathcal{O}_3 &=& \left( {Qt^c } \right)_\alpha ^\dag\; \sum_i g_i^2 {\rm Tr} [S_i  \Sigma S_i^\ast \Sigma^\ast  M_Q^\dagger P^\alpha]\,, 
\eeq
where the sum runs over the gauged generators of SU(4). The structure of the operators suggests that they may arise as one-loop corrections of the EW gauge bosons to the coupling of the elementary fermions to the dynamics, thus we expect the coefficients $C_{2,3}$ to be naively suppressed by a loop factor with respect to $C_1$. The expansion of the above operators can be conveniently expressed as:
\beq
\mathcal{O}_2 - \mathcal{O}_3 &=& (\mu_L - \mu_R) \left[ \frac{3 g^2 + {g'}^2}{16} \left( \sin (2\theta) + \frac{\cos (2\theta)}{\sqrt{2} f} h \right) + \right. \nonumber \\ 
& & + \left. i \frac{3 g^2 - {g'}^2}{16 \sqrt{2} f} \sin \theta\, \eta + \dots \right] (t_L t_R^c)^\dagger \,. \\
\mathcal{O}_2 + \mathcal{O}_3 &=& (\mu_L + \mu_R) \left[ \frac{3 g^2 - {g'}^2}{16} \left( \sin (2\theta) + \frac{\cos (2\theta)}{\sqrt{2} f} h \right) + \right. \nonumber \\
& & + \left. i \frac{3 g^2 + {g'}^2}{16 \sqrt{2} f} \sin \theta\, \eta + \dots \right] (t_L t_R^c)^\dagger \,.
\eeq
The first combination provides an example of linear coupling of $\eta$ which is proportional to the Sp(4) conserving part of the techni-quark mass $M_Q$, thus it cannot be set to zero. This proves that a linear coupling of $\eta$ to fermions is always generated in this model.
We also note that the linear coupling of $\eta$ would vanish if an exact SU(2)$_L$--SU(2)$_R$ symmetry were imposed on the model by gauging the full SU(2)$_R$ group: in this limit, the gauge coupling factor would be replaced by $3 g^2 - 3 {g'}^2 = 0$, as $g = g'$.

In our numerical analysis we will consider the most optimistic case as in Eq.~\eqref{eq:getaff}.
Using the tree-level coupling, the partial decay width in fermions is:
\beq
\Gamma (\eta \to f\bar{f}) = \frac{3 g_{\eta f \bar{f}}^2}{8 \pi} m_\eta \sqrt{1-\frac{4 m_f^2}{m_\eta^2}}\,.
\eeq

\subsection{Linear couplings to gauge bosons: anomalies}

The anomalous Wess-Zumino-Witten (WZW) term also breaks the $\eta$-parity, thus potentially generating linear couplings of the $\eta$ to two gauge bosons.
This coupling is very similar to the one mediating the decay of the $\pi^0 \to \gamma \gamma$ in QCD. It is associated with a triangle diagram of techni-quarks.
The anomaly diagram gives
    \begin{equation}
     \mathcal{M}=\frac{N_{\rm FCD} S}{16\pi^{2}\sqrt{2}f} \epsilon_{\mu\nu\rho\sigma}\epsilon^{\mu}_{I}(p_{1})
     \epsilon^{\nu}_{J}(p_{2})p_{1}^{\rho}p_{2}^{\sigma}\,,
    \end{equation}
with $\epsilon_{I,J}$ being the polarisations of the bosons $I,J$ and
    \begin{equation}
     S=\frac{1}{2} Tr[Y^{5}\{S^{I}, S^{J}\}]\,.
    \end{equation}
$Y^{5}$ is the $SU(4)$ operator associated with the $\eta$ particle and $S^{I/J}$ are the generators of SU(4) associated with the gauge bosons. 
$N_{\rm FCD}$ is the number of components of each techni-quark in the FCD group space, i.e. the dimension of the representation of $Q$ under the confining gauge symmetry: $N_{\rm FCD} = 2$ for SU(2)$_{\rm FCD}$ and $N_{\rm FCD} = 2 N$ for Sp(2N)$_{\rm FCD}$.
This amplitude corresponds to couplings of the form
\beq
i g_{\eta V_1 V_2} \eta \epsilon_{\mu \nu \alpha \beta} V_1^{\mu \nu} V_2^{\alpha \beta}\,,
\eeq
with (fixing $N_{\rm FCD} = 2$)
    \begin{gather}
     g_{\eta WW}=\frac{g^{2}s_{\theta}c_{\theta}}{16 \sqrt{2}\pi^{2}v}\,,
     \qquad
     g_{\eta ZZ}=\frac{(g^{2}-g'^{2})s_{\theta}c_{\theta}}{16 \sqrt{2}\pi^{2}v}\,,
     \qquad
     g_{\eta Z\gamma}=\frac{gg's_{\theta}c_{\theta}}{16 \sqrt{2}\pi^{2}v}\,, \nonumber \\
     g_{\eta\gamma\gamma}=0\,,
     \qquad
     g_{\eta g g}=0\,.
    \end{gather}
The coupling to gluons vanishes as the techni-quarks do not carry colour, while the coupling to photons vanishes due to the fact that U(1)$_{em}$ is fully embedded in SU(4).
Note also that no coupling of $h$ are generated by the WZW term, because $h$ is CP-even.

These couplings lead to the following partial decay widths:
  \begin{equation}
     \Gamma(\eta\rightarrow V_{1}V_{2})=\frac{g_{\eta V_{1}V_{2}}^{2}}{32\pi m_{\eta}^{3}}
     \left[\left(m_{\eta}^{2}-(m_{V_{1}}+m_{V_{2}})^{2}\right)
      \left(m_{\eta}^{2}-(m_{V_{1}}-m_{V_{2}})^{2} \right) \right]^{3/2} \frac{1}{1+\delta_{V_1 V_2}}\,,
    \end{equation}
where $\frac{1}{1+\delta_{V_1 V_2}}$ is a symmetry factor for identical final states.

The couplings  $\eta \to g g$ and $\eta \to \gamma \gamma$ will  be
generated by the top quark loop at the next leading order, after
taking into account the $\eta$-$t$-$t$ interaction discussed in the previous section. 
The corresponding partial decay widths are:
\beq \Gamma(\eta \rightarrow g g) &=& \frac{\alpha
\alpha_s^2m_\eta^3}{8\pi^2 m_W^2 \sin^2 \theta_W}\frac{g_{\eta tt
}^2 v^2}{m_t^2} F_1^2 (x_t)\,, \\ \nonumber  \\
\Gamma(\eta \rightarrow \gamma \gamma) &=& 1/2 ~
 N_c^2 \frac{\alpha^2}{\alpha_s^2}
\left(\frac{2}{3}\right)^4\Gamma(\eta \rightarrow g g)\,, \eeq
where $F_1(x_t)$ is the form factor for the top loop contribution
with $x_t = \frac{4 m_t^2}{m_\eta^2}$ being the rescaled top quark
mass squared, \beq F_1(x_t) = 1/2 ~ x_t \left(1+ (1-x_t) \sin^2
\left( x_t^{-1/2} \right)\right) \,, \eeq
and we neglect the light quarks contribution due to  their small
Yukawa interactions. It is  noted that in this  model, the
$W^{\pm}$ loop correction to the $\eta$-$\gamma$-$\gamma$
coupling vanishes due to the antisymmetric tensor
$\epsilon^{\mu \nu \rho \sigma}$, since there should be  no
further radiative corrections to the anomalous interaction.
These subleading couplings give negligible contribution to the width, however the coupling to gluons may play an important role for production at the LHC.

\subsection{Branching ratios}

 \begin{figure}[tb]
    \center
    \includegraphics[scale=1.]{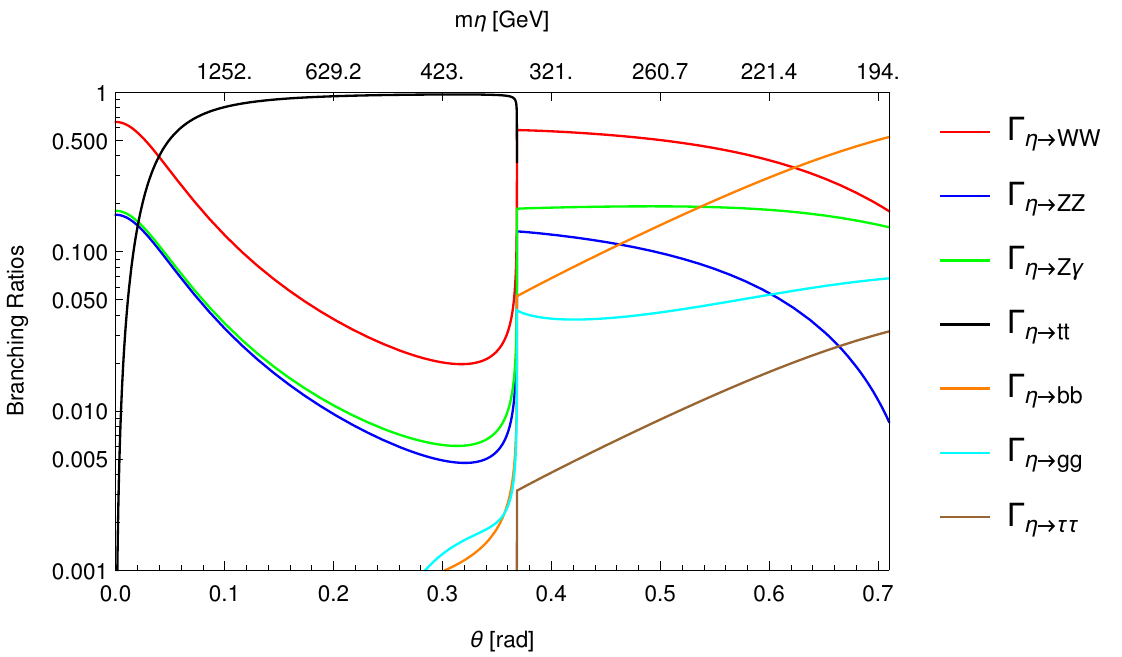}
    \caption{Branching ratios for the $\eta$ particle as a function of $\theta$ and $m_{\eta}$, for $m_{h}=125$ GeV. The angle $\theta$ range from $0$ to $\theta_{max, \rm{CMS}}$ defined in Section~\ref{sec:ewpt}.}
    \label{fig BR eta function of theta}
    \end{figure}

With the previous results, we can therefore calculate the
branching ratios for the $\eta$ particle. 
In the numerical examples, we will limit ourselves to the case of SU(2)$_{\rm FCD}$, and fix the coupling to fermions to the maximum value in Eq.(\ref{eq:getaff}).
We have two remaining free
parameters appearing in the couplings: $m_{\eta}$ and $\theta$. In the minimal case, they are related by a simple equation  $m_{\eta}=m_{h}/
\sin {\theta}$. In the following we will consider as an example the
limit case $\alpha \ll 1$, i.e. the case where the Higgs is almost
entirely composed of the $h$ particle. We set  $m_{h} \sim
m_{Higgs}\sim 125$~GeV and  compute the value of the branching
ratios as a function of $m_{\eta}$.
As illustrated by Fig. $\ref{fig BR eta function of theta}$, in the mass
region of $ 200~\mbox{GeV} < m_\eta <350~\mbox{GeV}$, the decay
width is dominated by the $W^+ W^-$, $Z \gamma$, $ZZ$ and $b \bar
b$ channels, and for $ m_\eta > 350~\mbox{GeV}$ $t \bar t$ becomes the
dominating final state. The branching ratio of $ \eta \to g g$ is
relatively large, which is  $6 \%$ for $\eta =200~ \mbox{GeV}$
and comparable to $\eta \to b \bar b$ in certain regions. However
for the decay $\eta \to \gamma \gamma$, its branching ratio is
negligible, accounting for $0.02$ percent in the low $m_\eta$  region due to the vanishing $W$ loop contribution.
In the figure we can also see that the bound on $\theta$ coming from the Higgs couplings and EWPTs, i.e. $\theta \leq 0.24$, restricts the $\eta$ to the heavy mass region, $m_\eta \geq 500$ GeV, which is above the $t\bar{t}$ threshold. We can therefore conclude that the scalar will always predominantly decay into tops, with the exception of very high masses (above $2.5$ TeV, i.e. $\theta < 0.05$) where the di-boson channels become prevalent.

\subsection{Production cross sections}

\begin{figure}[tb]
    \center
    \includegraphics[scale=1.0]{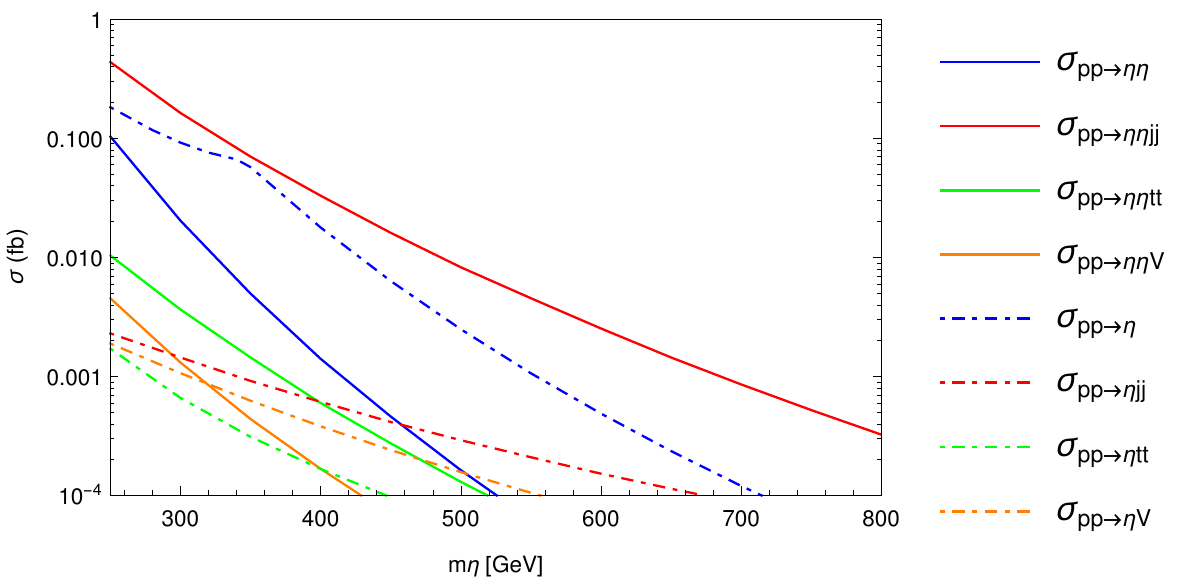}
    \caption{Production cross sections as a function of $m_\eta$ at the LHC Run II with $\sqrt s = 13$ TeV. We set the PDF to be MSTW2008NLO. The renormalization and factorization scales are fixed to be $\mu_R =\mu_F =\Sigma_f m_f/2$, where we sum over the massive final states. The cross sections of the pair  productions are drawn in solid lines while  the cross sections of corresponding single productions are drawn in dashing lines with matching colors.}
    \label{fig ProductionRate}
    \end{figure}

We continue to discuss the production mechanism of the $\eta$
particles at the LHC Run II. Since the linear couplings are loop suppressed, the pair production is generally
expected to be larger than the corresponding single production, but it does not apply to the gluon fusion processes due to the high momentum behavior of the off-shell form factor.
The  leading production channel is  $q q' \to \eta \eta + 2 j$, where we sum over two major effects.  One is the  direct vector bosons  fusion via  the $V$-$V$-$\eta$-$\eta$ vertex  and the other one is Higgs mediated VBF  pair fusion. The interference between these two diagrams turns out to be constructive  and the Higgs mediated VBF  pair fusion  will  play a comparable role in the large $m_\eta$ region. At a $\sqrt s = 13$ TeV LHC, with $m_\eta =250 ~\mbox{GeV}$, the cross section for $p p \to \eta \eta + 2 j$  is around $0.43$~fb (see Fig \ref{fig ProductionRate}).  The corresponding single process  is   $q q' \to \eta + 2j$, where there are  additional diagrams with $Z$-$\gamma$-$\eta$ interactions in addition to the  SM VBF diagrams, but its cross section could be  two orders smaller than the VBF pair production.  The next leading  production channels for the $\eta$ particle are  through the single and pair gluon fusions, i.e. $g g \to \eta $ and  $g g \to \eta \eta $, which are dominated by the top loop contribution. We can see from the simulation result that for  $m_\eta > 300 ~\mbox{GeV}$, $g g \to \eta$  is  more dominant due to the fact that  the momentum dependent form factor of $g g \to \eta \eta$  goes to zero in the large limit of $s = (p_{g1}+p_{g2})^2$.  Notice that there is  a transition at  $m_{\eta} = 2 m_t$ for the single gluon fusion process, which exactly reflects the fact that  the form factor from the top loop needs to  change at that point. In Fig. \ref{fig ProductionRate}, we also display the cross sections for the other major production channels, i.e. the associated production channels $ g g \to \eta \eta t \bar t$,  $g g \to \eta t \bar t $ and $q \bar q \to \eta V$, $q \bar q \to \eta \eta V$, where $V$ stands for all the SM gauge bosons $W,Z, \gamma$.  Those processes are  sensitive to the NLO QCD corrections and the choice of parton distribution functions (PDFs). In  the  MadGraph simulation  \cite{Alwall:2011uj}, we only take into account the leading order (LO) effects and ignore a sizable factor from the NLO correction.  We  choose the PDF set to be MSTW2008NLO \cite{Martin:2009iq}, while other PDFset choice could lead to $10\%$ variation for the hadronic cross sections.

The leading cross section is therefore always pair production: in the case with a significant coupling to fermions, this will result in a final state with four tops. However, the rates are very small, always below one fb, and they will also be dominated by the SM four top production which rates to about 15 fb at a centre of mass energy of 14 TeV~\cite{Bevilacqua:2012em}.
One possibility to reduce the SM background would be to tag the forward jets produced in association with the $\eta$'s, nevertheless this may be doable only with a very large integrated luminosity.

For completeness, we also calculated the production cross section at a linear electron-positron collider: the analytical results are shown in Eqs. (\ref{eq: ee
eta gamma}) and (\ref{eq: ee eta Z}). The main channel is the production in association with a neutral boson, i.e. $e^+ e^- \to \eta \gamma $ and $e^+ e^- \to \eta Z$:

 \beq
\sigma (e^+ e^- \to \eta \gamma) =
 \frac{\alpha ~ g_{\eta z \gamma}^2 \left(s- m_\eta^2 \right)^3}{s\left(\left(s- m_z^2\right)^2 + m_z^2 \Gamma_z^2\right)}
  \frac{ \left(\left(c_w^2- s_w^2\right)^2+4 s_w^4\right)}{12 c_w^2 s_w^2
  } \cdot \theta \left( s- m_\eta^2 \right)\,, \label{eq: ee eta gamma}
 \eeq
\beq \sigma(e^+ e^- \to \eta Z) &=&  \frac{ \alpha~ \left(\left(s-
{m_\eta}^2- {m_z}^2\right)^2-4 {m_\eta}^2 m_z^2\right)^{3/2} }{12
s^3 \left(\left(s- m_z^2\right)^2 + m_z^2 \Gamma_z^2\right)
{c_w}^2 {s_w}^2 }\cdot \mathcal{}\theta \left( s- \left(m_z+
m_\eta \right)^2 \right) \nonumber \\
&\cdot& \left( 8  g_{\eta z\gamma}^2 \left(s-m_z^2\right)^2 c_w^2
s_w^2 + g_{\eta z z}^2 s^2
\left(\left(c_w^2-s_w^2\right)^2+4 s_w^2\right) \right. \nonumber \\
&+&  \left. 4 g_{\eta z \gamma} g_{\eta zz} s \left(s-m_z^2\right)
c_w s_w \left(c_w^2-3 s_w^2\right) \right)\,. \label{eq: ee eta Z}
\eeq
\begin{figure}[tb!]
    \center
    \includegraphics[scale=1.0]{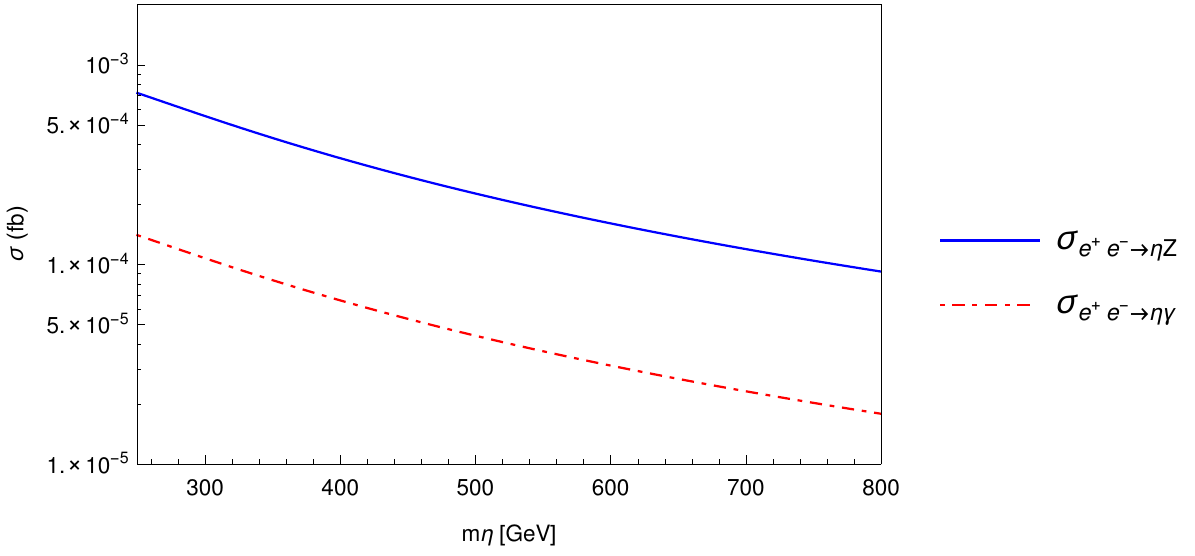}
    \caption{Production cross sections for the  $e^+ e^- \to \eta Z$ and $e^+ e^- \to \eta \gamma$ from the anomalous  $\eta$-$V$-$V'$ vertice in the  large  energy squared limit $s \gg m_\eta^2$. }
    \label{fig ProductionLinear}
    \end{figure}
In the case with only SM gauge bosons mediation, the cross
sections will increase with the center of mass energy squared
after reaching the final state threshold and then flatten to  a
constant in the high energy limit. In Fig. \ref{fig
ProductionLinear}, we plot the production cross sections for $e^+ e^- \to
\eta Z$ and $e^+ e^- \to \eta \gamma$ in the large limit $s \gg
m_\eta^2$.  Due to the small vertices of  $g_{\eta z \gamma }$ and
$g_{\eta z z}$, the cross sections turn out to be very small and
comparable to the relevant results at the hadronic collider.

\section{Constraints on the heavier Higgs boson} \label{sec:h2}

In the previous sections we have investigated the implications of the
Higgs boson measurements and of the electroweak precision
parameters on a general fundamental composite electroweak dynamics,
embracing both the ideas of a pNGB Higgs and of a composite
Higgs in terms of techni-fermions. However the simplest fundamental
dynamics we used as a guiding example, predicts the existence of
other composite states. As already discussed in the introduction,
preliminary lattice results indicate that the composite vector and
axial-vector states are expected rather heavy and outside the present
reach of the LHC (see for example \cite{Hietanen:2014xca}).
Concerning the scalar sector of the model, lattice results are still too
preliminary \cite{Arthur:2014lma}, but a first indication is that the scalar
composite $\sigma$ will be lighter. Therefore, the heavier mass eigenstate $h_2$ (see Eq.~\eqref{eq:mixing h sigma}) resulting from the mixing of the pNGB and techni-Higgs may well be within the reach of the LHC. In the effective Lagrangian description, this
second heavier Higgs $h_2$ can be characterised in terms of five parameters: the angles $\alpha$ and $\theta$, the (properly normalised) $\sigma$ couplings $\tilde{\kappa}_G$ and $\tilde{\kappa}_t$, and the mass $m_{h_2}$. 
The couplings of $h_2$ to SM gauge bosons $V = W^\pm, Z$, and the fermions (mainly the top) are given by:
\beq \label{eq:h2scaling}
\frac{g_{h_2 VV}}{g_{hVV}^{\rm SM}}& =& \sin (\theta - \alpha) + (\tilde{\kappa}_G -1 ) \sin \theta \cos \alpha\,, \nonumber \\ 
\frac{g_{h_2 f \bar{f}}}{g_{hf\bar{f}}^{\rm SM}}& =& \sin (\theta - \alpha) + (\tilde{\kappa}_t -1 ) \sin \theta \cos \alpha\,.
\eeq
To simplify the analysis we fix $\tilde{\kappa}_G$ and $ \tilde{\kappa}_t$, and show how the LHC can constrain the remaining parameters. 

\begin{figure}[tb!]
\center
\includegraphics[scale=0.7]{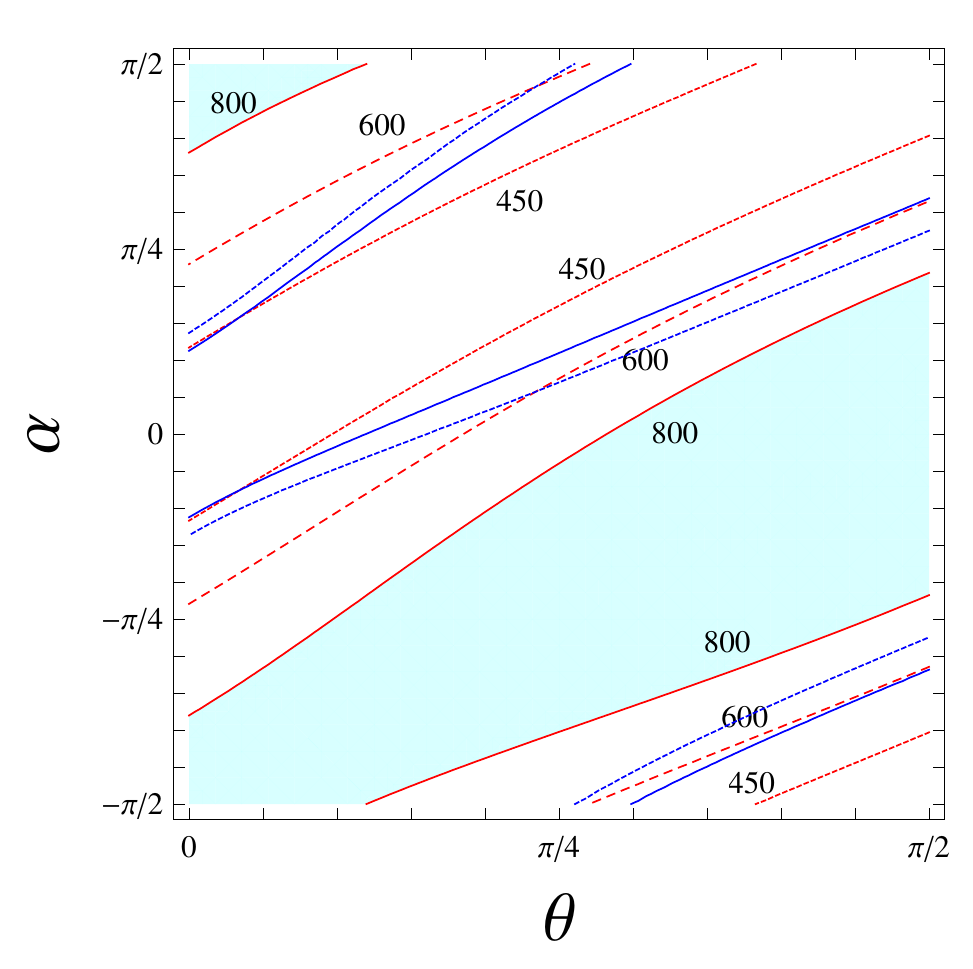}
\includegraphics[scale=0.6]{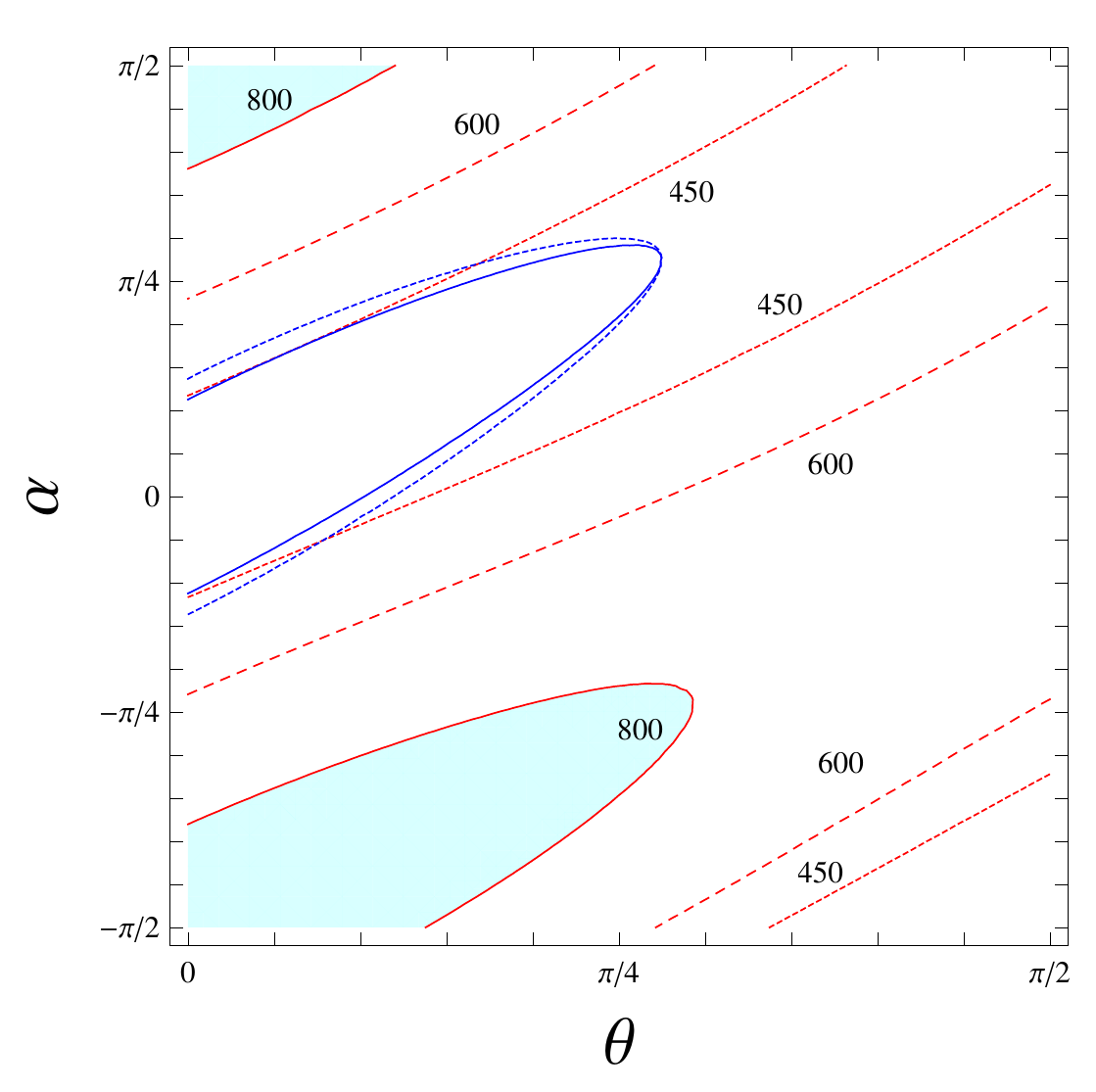}
\caption{95\% CL limit on the  $m_{h_2}$ from the CMS measurement of  $h_2 \to  ZZ \to 4l$. The red contours indicate the minimum value of  $m_{h_2}$ corresponding to  the upper limit of $\mu_\mathrm{exp}$; the blue contours are the bounds from the EWPT allowed region for  
$m_{h_2}= 450, 800$ GeV in dashing and solid lines respectively. In the left panel, we fix $\tilde{\kappa}_G=\tilde{\kappa}_t= 1.2$, while in the right panel $\tilde{\kappa}_G=1.0$ and $\tilde{\kappa}_t = 0.8$.} \label{fig ZZ4l}
\end{figure}

As $h_2$ has the same couplings of a SM Higgs, rescaled by the factors in the above formula, the constraints on its mass can be extracted by looking at searches of a Higgs at high mass, which is dominated by decays to $WW$ and $ZZ$.
In the following, we will use the CMS search for a heavy Higgs in the channel $h_2 \to ZZ \to 4l$ comprising the full Run 1 dataset \cite{Chatrchyan:2013mxa}, from which we  extract the 95\% confidence level limit on the signal strength $\mu$ as a function of the mass of $h_2$. This provides the most constraining search for large mass Higgses.  For $\tilde{\kappa}_G=\tilde{\kappa}_t$, there is only one rescaling factor between the $h_2$ coupling and the corresponding SM Higgs coupling in our model, thus $\mu_{th}$ is simply equal to the square of the scaling factor in Eq.(\ref{eq:h2scaling}).  Once the $\tilde{\kappa}$'s are fixed,  using the  fitted upper limit function of  $\mu_{exp} (m_{h_2})$,  for each point in the $(\alpha, \theta)$ plane a minimum value of $m_{h_2}$ can be extracted by requiring $\mu_{th} \leq \mu_{exp}$.  The result is shown in Figure~\ref{fig ZZ4l}. Note that the  oblique parameters depend weakly on the $h_2$ mass and tend to prefer the region $\alpha \sim \theta$ where the couplings of $h_1$ are close to SM-like, which is the region where $h_2$ tend to decouple. In the left panel of Fig.~\ref{fig ZZ4l} we show the case of $\tilde{\kappa}_G=\tilde{\kappa}_t = 1.2$. The region between the blue lines are allowed by the EWPTs, and it shrinks slightly with increasing mass of the heavy Higgs.  
In the preferred area, the lower bound on $h_2$ is always rather weak, being always below $600$ GeV.
Thus we can conclude that the second Higgs can still be fairly light and have escaped detection at the LHC.

In the case $\tilde{\kappa}_G \neq \tilde{\kappa}_t$, as there will be two different rescaling factors related to the vector boson couplings and SM fermion couplings, thus the signal strength is a more complicated function of $(\alpha, \theta, m_{h_2})$. Following the same procedure as above, we illustrate the bounds in the right panel of Figure~\ref{fig ZZ4l}, where $\tilde{\kappa}_G=1.0 $ and $\tilde{\kappa}_t = 0.8$. As before, in the region preferred by EWPTs, the lower bound on $m_{h_2}$ is rather loose, being always below $450$ GeV
This is quite general feature of this model therefore, as $h_2$ always tends to decouple if the lighter Higgs $h_1$ has couplings close to the SM ones.

\section{Conclusions} \label{sec:concl}

On the eve of the second run of the LHC experiments, it is of paramount importance to determine the status of the fundamental composite electroweak dynamics paradigm. We have therefore performed an analysis that  encompasses the simplest realisation bridging composite Goldstone Higgs models and Technicolor. By simplest we mean that it admits the most minimal fundamental realisation, also investigated via first principle lattice investigations. 

A typical model of fundamental composite dynamics will contain scalars behaving like pNGBs of the global symmetry breaking and are therefore light, scalars that are truly composite states, and spin-1 states.
Lattice data seem to indicate that the spin-1 states are always fairly heavy, above 2-3 TeV and with masses increasing for smaller $\theta$ parameters.
Thus, from this kind of scenario, the states that should be the first one to be studied at the LHC are scalars. 
In this work we focused on the minimal fundamental composite dynamics scenario, based on the symmetry breaking SU(4)/Sp(4). Together with a Higgs-like state, the pNGBs also include a singlet $\eta$.
We have investigated the interesting phenomenological interplay between the pNGB and techni-Higgs  interpretation of the discovered Higgs particle. These two states necessarily mix since they are both present in any fundamental four-dimensional realisation of a composite pNGB nature of the Higgs. Once the effective Lagrangian has been introduced and properly justified, we used the EWPTs as well as CMS and ATLAS most recent constraints on the Higgs couplings and decays  to constrain the effective coupling parameter space. 
We showed that a less fine tuned vacuum can be reached once a significant mixing between the two scalars is generated.
We then investigated the potential phenomenological impact of the singlet pNGB and the heavier Higgs-like state.
The $\eta$ cannot play the role of Dark Matter as it decays into gauge bosons via the WZW anomaly and to fermions via higher order operators (possibly generated by electroweak loops). However, the production rates at the LHC Run II, and at a future linear collider, are very small, making its detection very challenging.
The second Higgs may also show up in searches for Higgs-like states at high mass, and we showed that the present bounds are quite mild allowing for masses of a few hundred GeV.
This study can be considered as a benchmark for models of fundamental composite dynamics: non-minimal cases may contain more scalars with better 
detection prospects as they may be charged, while others may play the role of Dark Matter.

We have shown that the first LHC run is compatible with a composite nature of the Higgs mechanism in any of the limits considered, including Technicolor.  
Because of the link to the fundamental dynamics, we will be able, in the near future, to relate these constraints to direct first principle lattice simulations. 
Our results set the stage for the LHC Run II searches of natural composite dynamics at the Fermi scale.

\subsubsection*{Acknowledgements}
We wish to thank M. Gillioz for useful discussion, and J.B. Flament for providing us the updated code fitting the ellipses from signal strength measurements. AD is partially supported by Institut Universitaire de France. We also acknowledge partial support from
the Labex-LIO (Lyon Institute of Origins) under grant ANR-10-LABX-66 and FRAMA (FR3127, F\'ed\'eration de Recherche ``Andr\'e
Marie Amp\`ere") and the Danish National Research Foundation under the grant DNRF 90.


\end{document}